\documentclass[
  aps,
  prx,
  reprint,
  superscriptaddress,
  amsmath,
  amssymb,
]{revtex4-2}

\usepackage{iftex}
\usepackage{graphicx}
\usepackage{dcolumn}
\usepackage{bm,bbm}
\usepackage{physics}
\usepackage{algorithm}
\usepackage{algorithmicx}
\usepackage{algpseudocode}
\usepackage[colorlinks,
            linkcolor=blue,
            anchorcolor=blue,
            citecolor=blue,
            urlcolor=blue]{hyperref}

\ifPDFTeX
  \DeclareUnicodeCharacter{03B1}{\ensuremath{\alpha}}
  \DeclareUnicodeCharacter{03B2}{\ensuremath{\beta}}
  \DeclareUnicodeCharacter{03C3}{\ensuremath{\sigma}}
\fi

\graphicspath{{figures/}{../}{../data/}}

\newcommand{\Moller}{M{\o}ller}



\begin{document}

\title{Analytic Solution to Cumulative Coulomb Scattering in Boltzmann Theory}

\affiliation{Key Laboratory for Laser Plasmas (MoE) and School of Physics and Astronomy, Shanghai Jiao Tong University, Shanghai 200240, China}
\affiliation{Beijing National Laboratory for Condensed Matter Physics, Institute of Physics, Chinese Academy of Sciences, Beijing 100190, China}
\affiliation{School of Physics and Key Laboratory of Quantum State Construction and Manipulation (Ministry of Education), Renmin University of China, Beijing 100872, China}
\affiliation{School of Physical Sciences, University of Chinese Academy of Sciences, Beijing 100049, China}
\affiliation{DCI Joint research Team, Shanghai Jiao Tong University, Shanghai, 200240, China}
\affiliation{State Key Laboratory of Dark Matter Physics, Tsung-Dao Lee Institute, Shanghai Jiao Tong University, Shanghai 201210, China}

\author{Tie-Huai Zhang}
\affiliation{Key Laboratory for Laser Plasmas (MoE) and School of Physics and Astronomy, Shanghai Jiao Tong University, Shanghai 200240, China}
\affiliation{DCI Joint research Team, Shanghai Jiao Tong University, Shanghai, 200240, China}

\author{Wei-Min Wang}
\email{weiminwang1@sjtu.edu.cn}
\affiliation{Key Laboratory for Laser Plasmas (MoE) and School of Physics and Astronomy, Shanghai Jiao Tong University, Shanghai 200240, China}
\affiliation{DCI Joint research Team, Shanghai Jiao Tong University, Shanghai, 200240, China}

\author{Zi-Yan Zhang}
\affiliation{School of Physics and Key Laboratory of Quantum State Construction and Manipulation (Ministry of Education), Renmin University of China, Beijing 100872, China}
\affiliation{DCI Joint research Team, Shanghai Jiao Tong University, Shanghai, 200240, China}

\author{Yu-Tong Li}
\affiliation{Beijing National Laboratory for Condensed Matter Physics, Institute of Physics, Chinese Academy of Sciences, Beijing 100190, China}
\affiliation{School of Physical Sciences, University of Chinese Academy of Sciences, Beijing 100049, China}
\affiliation{DCI Joint research Team, Shanghai Jiao Tong University, Shanghai, 200240, China}

\author{Jie Zhang}
\email{jzhang1@sjtu.edu.cn}
\affiliation{Key Laboratory for Laser Plasmas (MoE) and School of Physics and Astronomy, Shanghai Jiao Tong University, Shanghai 200240, China}
\affiliation{DCI Joint research Team, Shanghai Jiao Tong University, Shanghai, 200240, China}
\affiliation{State Key Laboratory of Dark Matter Physics, Tsung-Dao Lee Institute, Shanghai Jiao Tong University, Shanghai 201210, China}

\date{\today}

\begin{abstract}

In plasmas, frequent soft Coulomb collisions gradually transfer energy and momentum, whereas a 
single close collision can change both abruptly and redirect a particle. This coexistence can 
shape runaway-electron avalanches in tokamaks and alpha-particle heating in inertial fusion, 
yet no first-principles analytic theory has determined how the cumulative statistics evolve 
under plasma conditions. Fokker--Planck and moment-matched models reproduce average transport 
but do not determine how scattering evolves from isolated events to cumulative relaxation across 
plasma conditions. Here we resolve this long-standing problem directly from Boltzmann theory. By 
asymptotically rescaling the exact angular expansion, we obtain the first quantitative map of 
the single-collision, core-tail coexistence, and low-order relaxation regimes within a unified 
Boltzmann framework. The rescaling also provides explicit analytic representations of the 
angular distribution across these regimes without empirical transition parameters. The theory 
further shows that small-angle models differ from the Boltzmann distribution even in the forward 
core, so appending a large-angle tail to a conventional core cannot recover the collision 
statistics. In fast-electron transport, this full-range redistribution leaves mean energy loss 
nearly unchanged yet counterintuitively enhances beam concentration and modifies the energy flux 
delivered to a finite region.

\end{abstract}

\maketitle

\section{Introduction}
\label{sec:introduction}

In plasmas, soft Coulomb collisions dominate particle thermalization and mean
energy transfer~\cite{rosenbluth_fokker-planck_1957,Landau_transport_1965},
whereas a single close collision can abruptly change a particle's energy,
momentum, and direction. The coexistence of these outcomes matters whenever a
plasma process depends not only on mean energy loss but also on particle
trajectories or on rare collisions that trigger a threshold response.
Representative examples include the collimation and localized heating of fast
electrons~\cite{robinson_theory_2014}, runaway-electron avalanche multiplication
by \Moller{} knock-on collisions~\cite{rosenbluth_theory_1997,breizman_physics_2019},
and alpha-particle energy deposition and suprathermal-ion production in fusion
fuel~\cite{fisher_fast_1994,hartouni_evidence_2023}. These phenomena expose a
broader unresolved question: under given plasma conditions, does cumulative Boltzmann 
scattering retain single-collision statistics, develop core-tail coexistence, or 
reduce to low-order relaxation?

\begin{figure}[!ht]
  \centering
  \includegraphics[width=\columnwidth]{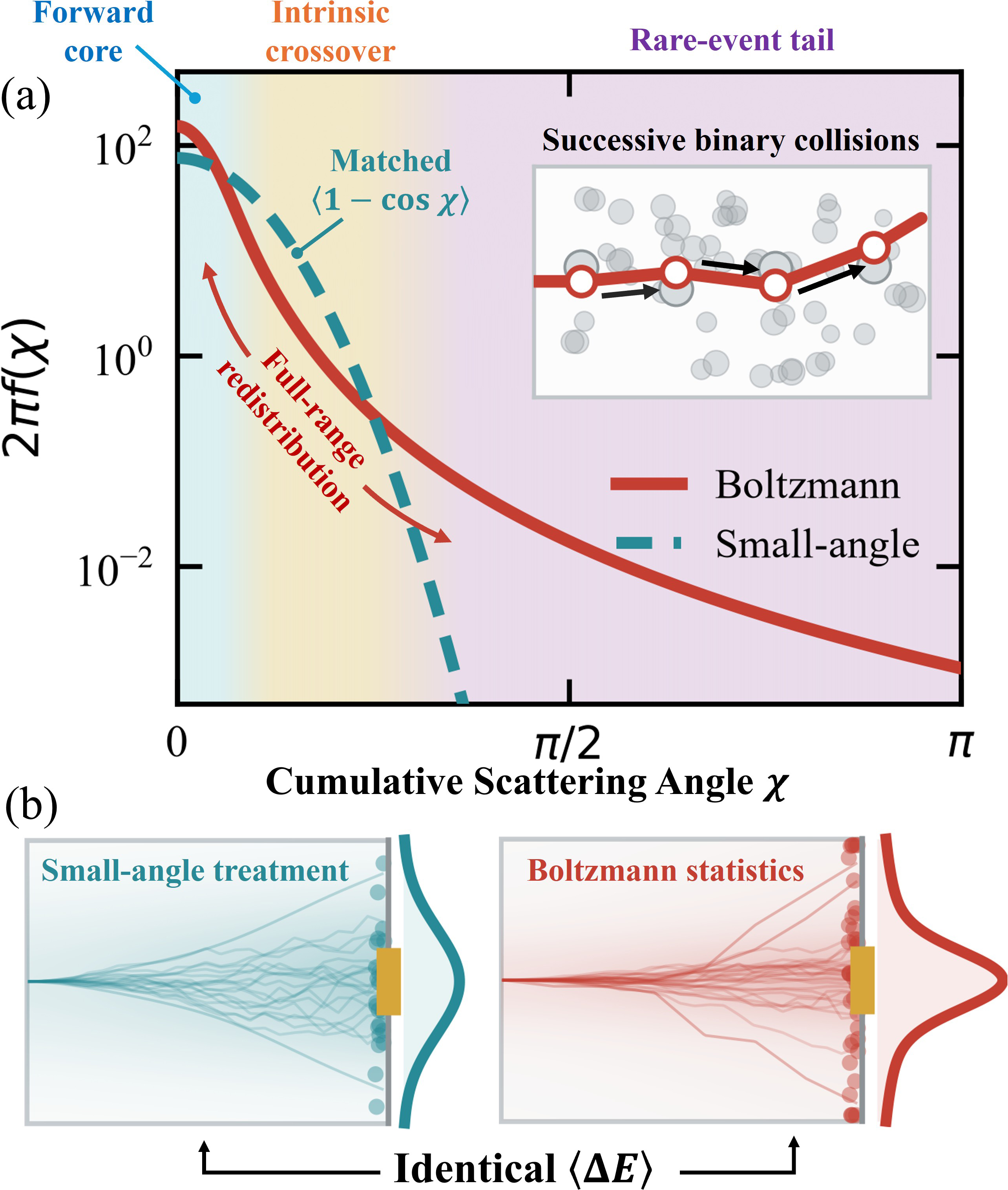}
  \caption{Conceptual overview. (a) The small-angle distribution is explicitly
  matched to $\langle1-\cos\chi\rangle$, whereas the Boltzmann description
  is obtained directly from the binary cross section. Despite the same first
  moment, statistical behaivors differs across the forward core, intrinsic crossover,
  and rare-event tail. The inset illustrates successive binary collisions 
  experienced by a test particle in a short time interval $\Delta t$.
  (b) With identical mean energy loss, the two treatments would yield different beam
  concentration. The side curves show transverse energy-flux profiles, and the
  gold bars mark the finite central collection region. The beam becomes 
  counterintuitively more concentrated after applying the full-angle Boltzmann 
  statistics.}
  \label{fig:conceptual-overview}
\end{figure}

To make this problem tractable, plasma models generally adopt a statistical form 
associated with a selected limit. Frequent weak collisions are represented by a
Fokker--Planck collision operator or by a distribution constrained to reproduce
mean transport~\cite{nanbu_theory_1997,nanbu_weighted_1998}, whereas large-angle
events drawn from the single-collision cross section are appended
explicitly~\cite{turrell_self-consistent_2015,higginson_full-angle_2017,
embreus_relativistic_2018}. These strategies are effective in their intended limits, 
but they do not provide a unified, first-principles determination of whether given 
conditions produce single-collision, core-tail coexistence, or low-order relaxation 
statistics, nor where the boundaries between these regimes lie. A second limitation concerns 
the probability distribution within a given regime. Splicing can recover a power-law 
tail but does not generally preserve the correct mean momentum and energy transfer 
without an additional moment constraint~\cite{ray_angus_monte-carlo_2026}. Even when
the first transport moment is enforced, it fixes the mean transfer but not the
probabilities of forward, intermediate, or rare large-angle deflections.
Figure~\ref{fig:conceptual-overview}(a) illustrates how a moment-constrained description 
can therefore differ from the Boltzmann result throughout the angular distribution. 
In Boltzmann theory, the statistical regime, angular distribution, and transport moments 
all emerge from the same binary collision process. Therefore the full distribution cannot 
be reconstructed by appending a rare-event tail to an assumed Fokker--Planck core. Nearly 
identical mean energy loss can consequently coexist with different trajectories, spatial
transport, and threshold-crossing probabilities, as sketched in 
Fig.~\ref{fig:conceptual-overview}(b). Transport equivalence does not imply
distributional equivalence. The unresolved problem is thus twofold: to determine which 
statistical regime arises under given conditions and to derive its distribution without 
imposed splicing or moment constraint.

The difficulty originates in the strong forward-scattering singularity of the
bare Coulomb cross section. Plasma screening regularizes the collision integral
but leaves the binary cross section concentrated within an angular scale of order
$\Lambda^{-1}$, where $\ln\Lambda$ is the Coulomb logarithm and $\Lambda$
quantifies the separation between the screening and close-collision
scales~\cite{krommes_introduction_2019}. The formally exact distribution of the 
cumulative scattering angle $\chi$ has been known since 1940 through the 
Goudsmit--Saunderson (GS) Legendre series
\cite{goudsmit_multiple_1940,goudsmit_multiple_1940-1}, a representation
rooted in Boltzmann elastic-collision dynamics~\cite{bobylev_theory_2000}.
Resolving this distribution requires Legendre orders $l\sim\Lambda$. For Coulomb
logarithms typical of laboratory and astrophysical plasmas,
$\ln\Lambda\simeq5\!-35$, the required range extends from $l\sim10^2$ to
$10^{15}$. Direct summation is therefore impractical across most of the plasma
regime. More importantly, the formal series reveals neither how the statistical 
regimes and their boundaries emerge nor how the angular distribution is organized 
within each regime. Extracting this structure directly from the 
screened GS coefficient hierarchy at plasma-relevant cutoff angles has remained 
analytically unresolved for decades.

Landau derived the grazing-collision limit from small deflections in
individual collisions~\cite{Landau_transport_1965}, while Moli\`ere developed a
systematic small-angle expansion~\cite{bethe_molieres_1953}. Lewis retained the
GS framework to obtain angular and spatial transport moments
~\cite{lewis_multiple_1950}, and Bobylev and Nanbu established its operator
connection to the Landau limit~\cite{bobylev_theory_2000}. Together, these works
establish the limiting physics, transport properties, and operator structure,
but do not determine how cumulative Boltzmann scattering moves between statistical 
regimes or provide analytic distributions across them under plasma conditions.
Later numerical GS methods enabled the evaluation and sampling of scattering 
probabilities in condensed-matter particle transport through variable 
transformations, high-order summation, and pretabulation 
\cite{kawrakow_electron_1996,kawrakow_representation_1998,
kadri_incorporation_2009,novak_new_2024}. These methods are practical in
regimes where the cutoff angle is comparatively large and the required
Legendre order remains manageable.
They make the distribution numerically accessible but do not expose its analytic 
regime structure, and they are inapplicable at the much smaller cutoff angles 
typical of plasmas \cite{higginson_full-angle_2017}.

We close both gaps by analytically resolving the screened GS coefficient hierarchy 
for a general axisymmetric elastic cross section with a power-law small-angle singularity.
An asymptotic representation with controlled remainders reduces the Coulomb
coefficients to a common rescaled function. This reduction yields, for the first time, a 
quantitative map of the single-collision, core-tail coexistence, and low-order relaxation 
regimes within one Boltzmann framework and determines their domains of validity. It also 
provides analytic representations of the angular distribution across these regimes and fixes 
the intrinsic crossover in the core-tail coexistence regime without empirical transition 
parameters. Since the derivation is not restricted to a particular binary cross
section, the framework extends beyond Rutherford scattering to other elastic
processes, such as \Moller{} scattering between identical electrons. A central
consequence is that the Boltzmann distribution differs from small-angle
descriptions in the forward core as well as the rare-event tail. We demonstrate
this full-range difference in fast-electron transport, where probability shifts
from intermediate deflections toward a sharper core while finite-angle events
persist. Despite nearly identical mean energy loss, the beam becomes
counterintuitively more concentrated, as presented in Fig. \ref{fig:conceptual-overview}(b). 
Across the propagation distances and  collection apertures examined, the resulting enhancement 
of the energy flux delivered to the central region reaches about \(10\%\) 
(Sec. \ref{sec:electron-transport}).

\section{Analytic Structure of Cumulative Boltzmann Scattering}
\label{sec:analytic-structure}


\subsection{Analytic rescaling of the Legendre coefficients}
\label{sec:rescaled-spectrum}

Consider a sufficiently short interval $\Delta t$ over which the relative
energy and local plasma conditions can be treated as fixed. Binary collision
events then occur independently at the rate $\sigma_t n v_r$, where $\sigma_t$
is the total cross section, $n$ is the background density, and $v_r$ is the
relative speed. The number of collisions experienced by a particle therefore
follows a Poisson distribution with mean $N=\sigma_t n v_r\Delta t$.
Goudsmit--Saunderson theory sums over these successive events and gives the
cumulative angular distribution~\cite{goudsmit_multiple_1940,goudsmit_multiple_1940-1}
\begin{equation}
\begin{aligned}
  f_{\rm GS}(\chi)
  &=\sum_{l=0}^{\infty}\frac{2l+1}{4\pi}
    e^{-s_l}P_l(\cos\chi),\\
  s_l&=N\left(1-\tilde{\sigma}_l\right).
\end{aligned}
  \label{eq:GS-distribution}
\end{equation}
Here $\chi$ is the net deflection accumulated during $\Delta t$, and
$e^{-s_l}=\langle P_l(\cos\chi)\rangle$ is the $l$th Legendre coefficient of
the cumulative distribution. Its relation to the binary collision physics is
direct. For an axisymmetric cross section
$\sigma_{\rm bi}(\theta)=\dd\sigma/\dd\Omega$, where $\theta$ is the deflection
in one collision, the normalized single-collision coefficient is
\begin{equation}
  \tilde{\sigma}_l
  =\frac{2\pi}{\sigma_t}
   \int_{\theta_m}^{\theta_M}\sigma_{\rm bi}(\theta)
   P_l(\cos\theta)\sin\theta\,\dd\theta
  =\left\langle P_l(\cos\theta)\right\rangle,
  \label{eq:single-legendre}
\end{equation}
Here $\theta_m$ is the cutoff angle set by screening, and $\theta_M$ is the
largest nonredundant scattering angle. For independent collisions, successive
Legendre coefficients $\tilde{\sigma}_l$ multiply, while averaging over the
Poisson-distributed collision count produces the exponential factor $e^{-s_l}$
in Eq.~\eqref{eq:GS-distribution}. Bobylev and Nanbu subsequently showed that this
GS construction is generated by the Boltzmann collision operator in the
short-time limit~\cite{bobylev_theory_2000}. Under the assumptions stated above,
Eqs.~\eqref{eq:GS-distribution} and~\eqref{eq:single-legendre} give the short-time
angular statistics generated by Boltzmann elastic collisions. The $l=0$
coefficient enforces normalization, the first few coefficients determine
low-order transport, and the full sequence $\{e^{-s_l}\}$ uniquely determines
the angular distribution. The complete binary-collision coefficient sequence,
rather than any selected transport moment, is therefore required.

For screened Coulomb scattering, $\theta_m\simeq2/\Lambda$, where
$\ln\Lambda=\ln\cot(\theta_m/2)$. Increasing $\Lambda$ narrows the binary cross
section around the forward direction. Low-order Legendre polynomials cannot
resolve how the cross section varies within this narrow angular region. This
information enters through its high-order coefficients. We therefore examine
the large-$l$ behavior without smoothing the small-angle singularity. For
$l\gg1$,
the uniform approximation naturally introduces $\rho=l+1/2$ through
\begin{equation}
  P_l(\cos\theta)
  \simeq\sqrt{\frac{\theta}{\sin\theta}}J_0(\rho\theta)
  \label{eq:legendre-bessel}
\end{equation}
This standard Legendre-to-Bessel approximation underlies classical
multiple-scattering analyses~\cite{bethe_molieres_1953,
winterbon_finite_angle_1987,kawrakow_representation_1998}.
Substituting Eq.~\eqref{eq:legendre-bessel} into
Eq.~\eqref{eq:single-legendre} converts the high-order coefficient into a
Bessel integral. Because the cross section is concentrated at small angles,
this integral is controlled by its local small-angle asymptotics. For
a general elastic kernel with
$\sigma_{\rm bi}(\theta)\sim\theta^{-p}$ as $\theta\to0$, the normalized angular
measure has the Laurent expansion
\begin{equation}
  \sqrt{\frac{\theta}{\sin\theta}}
  \frac{2\pi\sigma_{\rm bi}(\theta)\sin\theta}{\sigma_t}
  =\frac{1}{\mathcal A}
   \sum_{j=0}^{\infty}c_{-p+1+j}\theta^{-p+1+j}.
  \label{eq:laurent-kernel}
\end{equation}
Here $\mathcal A^{-1}$ is the amplitude of the leading $\theta^{-p+1}$ term
after normalization by $\sigma_t$, and $c_{-p+1}=1$. The remaining $c_j$
retain the subleading angular structure of the binary cross section.
Combining Eqs.~\eqref{eq:single-legendre}--\eqref{eq:laurent-kernel} gives the
controlled representation
\begin{equation}
  \tilde{\sigma}_l=
  \frac{1}{\mathcal A}\sum_{j=0}^{K}c_{-p+1+j}
  \int_{\theta_m}^{\theta_M}J_0(\rho\theta)
  \theta^{-p+1+j}\,\dd\theta+R_l,
  \label{eq:general-asymptotic-coefficient}
\end{equation}
Here $R_l$ collects the error of the Legendre--Bessel approximation
and the remainder of the truncated Laurent expansion. The analytic evaluation
and the asymptotic properties used below are given in
Appx.~\ref{app:spectral-expansion}. Equation~\eqref{eq:general-asymptotic-coefficient}
shows that the pole order fixes the large-$l$ scaling, while the coefficients
$c_j$ retain the collision-specific angular dependence.

Charged-particle scattering belongs to the Coulomb class $p=4$. For the
Rutherford and spin-averaged \Moller{} kernels, all even Laurent coefficients
vanish, so Eq.~\eqref{eq:general-asymptotic-coefficient} reduces to integrals of
$\theta^{2k-3}J_0(\rho\theta)$. Each integral has a closed form in bounded
combinations of generalized hypergeometric functions. With
$\vartheta_M=\rho\theta_M/2$ and $\lambda=\theta_M/\theta_m$, the result can be
written compactly by setting $\tilde{\sigma}_l\simeq\varsigma(\rho)$:
\begin{equation}
\begin{split}
  \mathcal A\,\varsigma(\rho)
  ={}&\frac{c_{-3}}{2\theta_m^2}
      \mathcal I_{-3}(\lambda,\vartheta_M)
     +\frac{c_{-1}}{2}\mathcal I_{-1}(\lambda,\vartheta_M)\\
    &+\frac{1}{2}\sum_{k=0}^{\infty}c_{2k+1}\theta_m^{2k+2}
      \mathcal I_{2k+1}(\lambda,\vartheta_M).
  \label{eq:bounded-coefficient}
\end{split}
\end{equation}
The functions $\mathcal I_j$ remain bounded for $\lambda>1$. Their definitions
and asymptotic limits are given in Appx.~\ref{app:spectral-expansion}.
Unlike a high-order recurrence or a direct Legendre integral,
Eq.~\eqref{eq:bounded-coefficient} remains well conditioned as the cutoff angle
decreases.
As a computational corollary, Eq.~\eqref{eq:bounded-coefficient} provides a
stable and accurate basis for pretabulating the GS coefficients used in Monte
Carlo transport codes, without repeatedly evaluating highly oscillatory Legendre
integrals.

\begin{figure}[htbp]
  \centering
  \includegraphics[width=\columnwidth]{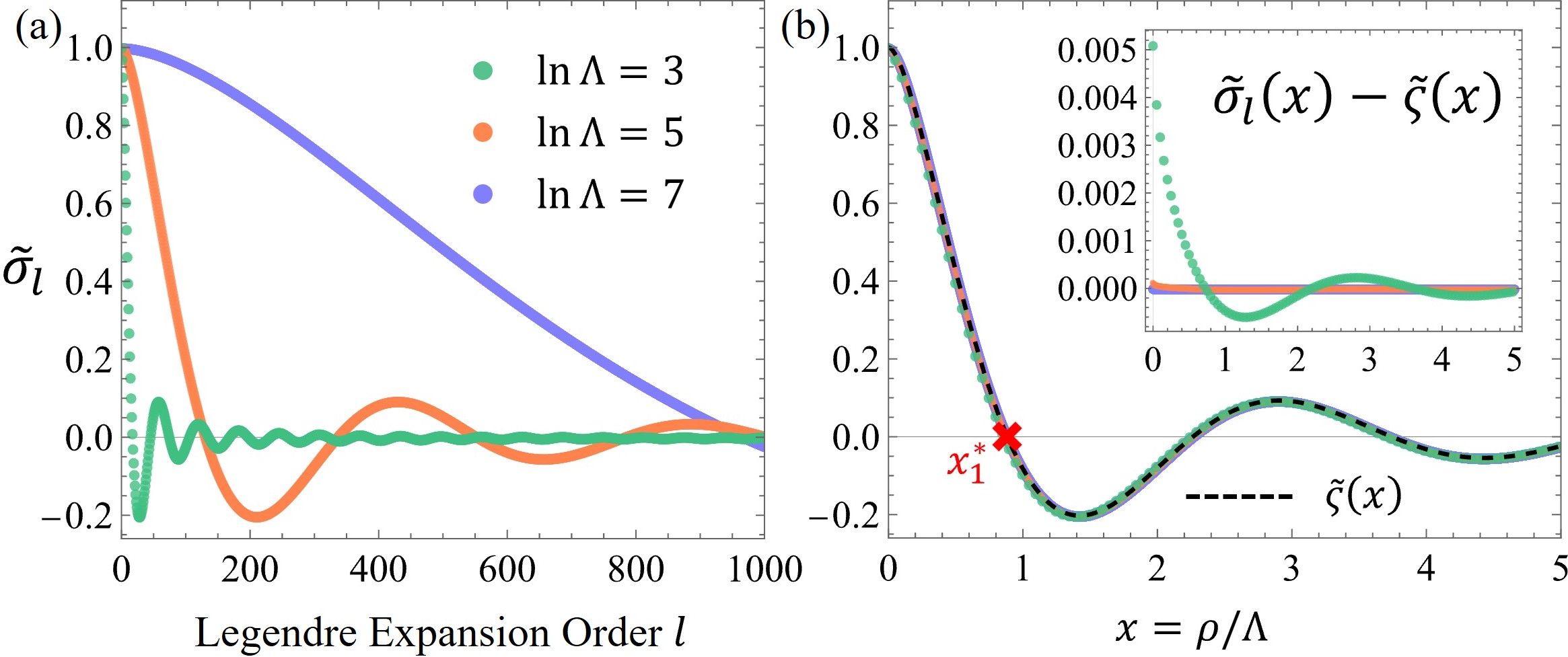}
  \caption{Rescaling of the Coulomb Legendre coefficients.
  (a) Normalized coefficients $\tilde{\sigma}_l$, evaluated directly from the
  Legendre integral for $0\leq l\leq1000$, at $\ln\Lambda=3$ (green), $5$
  (orange), and $7$ (purple). The variation of the coefficients shifts to progressively
  higher $l$ as the Coulomb logarithm increases. (b) Directly evaluated
  coefficients for the same three Coulomb logarithms, plotted against
  $x=\rho/\Lambda$ with $\rho=l+1/2$. All three data sets collapse onto the
  $\Lambda\rightarrow\infty$ rescaled coefficient function
  $\tilde{\varsigma}(x)$ obtained from the bounded hypergeometric
  representation (black dashed line). The red cross marks its first zero
  $x_1^*\simeq0.877$. The inset shows the signed residual
  $\tilde{\sigma}_l(x)-\tilde{\varsigma}(x)$.}
  \label{fig:rescaled-spectrum}
\end{figure}

For Rutherford scattering,
$\mathcal A=\tfrac18\cot^2(\theta_m/2)=\Lambda^2/8$. When $\Lambda\gg1$, the
$\mathcal I_{-3}$ term dominates Eq.~\eqref{eq:bounded-coefficient}, while all
remaining terms are smaller by $O(\Lambda^{-2}\ln\Lambda)$. Moreover, the
dependence on $l$ and $\Lambda$ within $\mathcal I_{-3}$ is captured by
rescaling the Legendre order with $\Lambda$:
\begin{equation}
  x\equiv\frac{\vartheta_M}{\lambda}
  =\frac{\rho\theta_m}{2}\simeq\frac{\rho}{\Lambda}.
  \label{eq:rescaled-variable}
\end{equation}
This rescaling yields
\begin{equation}
  \tilde{\sigma}_l
  =\tilde{\varsigma}(x)
   +O\!\left(\Lambda^{-2}\ln\Lambda\right)
   +R_l,
  \label{eq:rescaled-spectrum}
\end{equation}
where the leading function $\tilde{\varsigma}$ is independent of the
Coulomb logarithm. In the nonoscillatory interval it has the elementary form
\begin{equation}
  \tilde{\varsigma}(x)
  \simeq 1+2x^2\ln x-2(1-\gamma_E)x^2-\frac{x^4}{4},
  \qquad x\lesssim x_1^*,
  \label{eq:rescaled-core-branch}
\end{equation}
where $\gamma_E$ is the Euler--Mascheroni constant and
$x_1^*\simeq0.877$ is the first zero of $\tilde{\varsigma}$. Beyond this zero,
$\tilde{\varsigma}$ oscillates and decays as
\begin{equation}
  \tilde{\varsigma}(x)
  =\frac{\cos(2x+\pi/4)}{\sqrt{\pi}\,x^{3/2}}
   +O\!\left(x^{-5/2}\right).
  \label{eq:rescaled-oscillatory}
\end{equation}
Figure~\ref{fig:rescaled-spectrum} tests this rescaling
directly against the defining Legendre integral, without fitting parameters.
Before rescaling, the three coefficient sequences in
Fig.~\ref{fig:rescaled-spectrum}(a) vary over widely separated ranges of $l$:
increasing $\ln\Lambda$ pushes both the first zero and the subsequent
oscillations to progressively higher order.
After the rescaling $x=\rho/\Lambda$, however, all numerical coefficients
collapse onto the same curve [Fig.~\ref{fig:rescaled-spectrum}(b)]. The
collapse persists through the first zero, marked by the red cross at $x_1^*$,
and across the sign-changing oscillatory sector up to $x=5$. It therefore
tests the rescaling beyond the initial nonoscillatory branch. The
collapsed data coincide with the parameter-free hypergeometric
representation of $\tilde{\varsigma}(x)$ when $\Lambda\rightarrow\infty$.

The inset in Fig. \ref{fig:rescaled-spectrum}(b) resolves the signed 
finite-$\Lambda$ residual. Its largest magnitude
occurs near the low-order end and remains below $5\times10^{-3}$ even for
$\ln\Lambda=3$, and decreases rapidly as $\ln\Lambda$ increases. This behavior
is consistent with the correction structure in
Eq.~\eqref{eq:rescaled-spectrum} and shows that the rescaled curve
is already quantitatively accurate well before the large Coulomb logarithms
typical of plasmas are reached.

Through Eq.~\eqref{eq:rescaled-variable}, the first zero $x_1^*$ corresponds
to $l+1/2\simeq x_1^*\Lambda$. Resolving the binary-collision coefficients
through this first zero therefore requires $l=O(\Lambda)$, corresponding to
$l\sim10^2$--$10^{15}$ for typical plasma Coulomb logarithms. Equations
~\eqref{eq:rescaled-spectrum}--\eqref{eq:rescaled-oscillatory} replace these
plasma-dependent high-order calculations by one function of
$x=\rho/\Lambda$, with controlled $O(\Lambda^{-2}\ln\Lambda)$ corrections.
The rescaled form also exposes information discarded by a small-angle closure.
In particular, the
nonanalytic $x^2\ln x$ term in Eq.~\eqref{eq:rescaled-core-branch} carries the
signature of the $\theta^{-4}$ binary singularity. The next subsection shows
how that nonanalytic term shapes the rare-event asymptotics while the same
coefficient function determines the forward scattering distribution.

\subsection{Joint emergence of\\ the forward scattering core and rare-event tail}
\label{sec:core-tail-emergence}

We now invert the rescaled Legendre expansion to expose how different angular
populations emerge. The limiting value $e^{-s_l}\to e^{-N}$ as $l\to\infty$
is the probability that no collision occurs during $\Delta t$. Separating this
singular contribution from the smooth scattered population gives
\begin{equation}
  f_{\rm GS}(\chi)=e^{-N}\delta_\Omega(\chi)+g(\chi),
  \qquad \int\delta_\Omega(\chi)\,\dd\Omega=1,
  \label{eq:GS-smooth-decomposition}
\end{equation}
where $g$ contains all histories with at least one collision. Applying the
Legendre--Bessel approximation to the remaining smooth sum in Eq.~\eqref{eq:GS-distribution} 
and using Eq.~\eqref{eq:rescaled-spectrum} yields
\begin{equation}
\begin{aligned}
  &g(\chi)\simeq
  \sqrt{\frac{\chi}{\sin\chi}}\frac{\Lambda^2}{2\pi}
  \int_0^\infty x H(x)J_0(kx)\,\dd x,\\
  &H(x)=e^{N[\tilde{\varsigma}(x)-1]}-e^{-N},
  \qquad k=\Lambda\chi.
\end{aligned}
  \label{eq:GS-hankel-kernel}
\end{equation}
Equation \eqref{eq:GS-hankel-kernel} is the central angular representation used below.
With the dependence of $H$ on $N$ left implicit, it contains no prescribed
division between weak and strong collisions, and the two angular limits follow
from different asymptotic regions of the same integrand.

We first consider the forward population. In the multiple-collision range its
natural angular scale is the Moli\`ere width~\cite{bethe_molieres_1953}
\begin{equation}
  \omega^2=\frac{2NB}{\Lambda^2},
  \qquad B=-W_{-1}\!\left[-e^{-2(1-\gamma_E)}N^{-1}\right].
  \label{eq:moliere-width}
\end{equation}
Here $W_{-1}$ is the lower real branch of the Lambert function and
$B\simeq\ln(N'\ln N')$ for $N'=e^{2(1-\gamma_E)}N\gg1$. This scale follows from the logarithmic
$x^2\ln x$ term in Eq.~\eqref{eq:rescaled-core-branch}, and is not obtained by
fitting the final distribution. To resolve the shape around this scale, set
$z=\omega^2\rho^2$ and expand the kernel in the Laguerre functions
$\ell_n(z)=e^{-z/2}L_n(z)$. These functions transform term by term under the
zeroth-order Hankel integral~\cite{hardy_summation_1932,szego_orthogonal_1975},
giving the analytic core representation
\begin{equation}
  g_{\rm core}(\chi)=
  \sqrt{\frac{\chi}{\sin\chi}}\frac{1}{2\pi\omega^2}
  \sum_{n=0}^{\infty}(-1)^n h_n
  \ell_n\!\left(\frac{\chi^2}{\omega^2}\right).
  \label{eq:laguerre-core}
\end{equation}
Here
$h_n=\int_0^\infty H[\sqrt{z}/(\omega\Lambda)]\ell_n(z)\,\dd z$ are the
Laguerre moments of the cumulative kernel $H$, and their asymptotic evaluation for
particle sampling is given in the Supplemental Material (SM) Sec. I~\cite{SM}. 
The $n=0$ term produces the familiar Gaussian factor
$\exp[-\chi^2/(2\omega^2)]$, whereas the higher terms determine the peak and
intermediate-angle shoulders. These terms are projections of the same GS
expansion that determines $\omega$, not empirical corrections to an assumed
Gaussian. Fixing $\langle P_1(\cos\chi)\rangle=e^{-s_1}$ preserves the mean
momentum transfer but leaves these higher components unconstrained. Equal mean
momentum and energy transfer can therefore accompany different forward peaks
and intermediate-angle populations.

The rare-event tail follows from the opposite asymptotic reading of
Eq.~\eqref{eq:GS-hankel-kernel}. At $\chi\gg\omega$, and hence $k\gg1$, the
rapid oscillations of $J_0(kx)$ isolate the behavior of $H$ near $x=0$.
Equation~\eqref{eq:rescaled-core-branch} gives
\begin{equation}
  H(x)=1-e^{-N}+2Nx^2\ln x+O(Nx^2)
  \qquad (x\to0).
  \label{eq:hankel-kernel-origin}
\end{equation}
The constant does not contribute to the large-$k$ expansion for $k>0$. 
The first surviving term is instead the nonanalytic $x^2\ln x$ 
contribution inherited directly from the $\theta^{-4}$ singularity
of the binary cross section. Applying the asymptotic theorem for Hankel
integrals with logarithmic singularities~\cite{wong_asymptotic_1977} gives
\begin{equation}
\begin{aligned}
  g_{\rm tail}(\chi)&\simeq
  \sqrt{\frac{\chi}{\sin\chi}}
  \frac{4N\Lambda^2}{\pi k^4}\\
  &\quad\times\left[1+\frac{32N}{k^2}
  \left(\ln\frac{k}{2}-\frac12\right)
  +o\!\left(\frac{N}{k^2}\right)\right].
  \label{eq:analytic-tail}
\end{aligned}
\end{equation}
The leading term is the $\chi^{-4}$ law expected when one close collision
dominates a rare large deflection. The next $\chi^{-6}\ln\chi$ term is already
a cumulative effect: its nonlinear dependence on $N$ accounts for the weak
collisions accompanying that event. It has no counterpart in a construction
that simply appends the binary cross section as a large-angle tail, and becomes
quantitatively important before the leading power law is fully established.

\begin{figure}[t]
  \centering
  \includegraphics[width=\columnwidth]{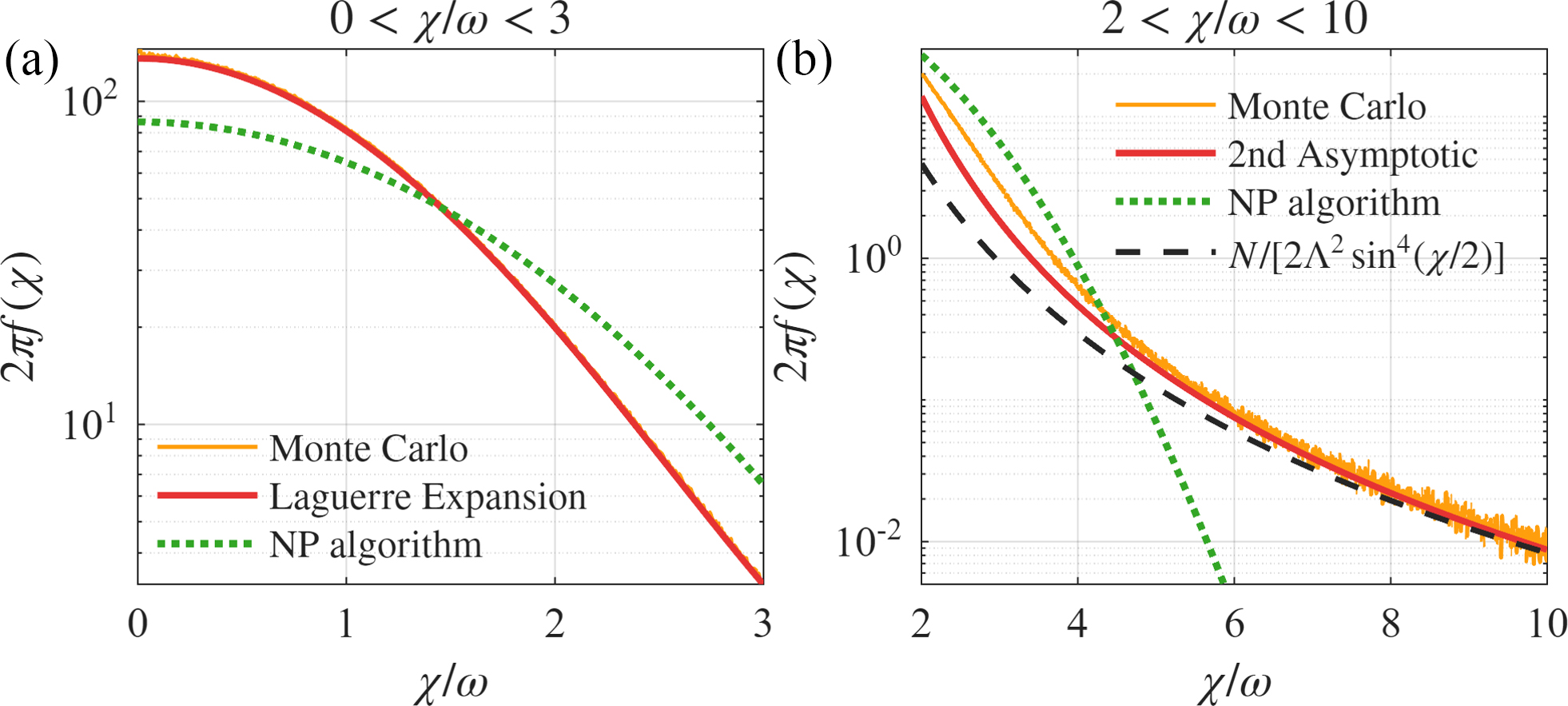}
  \caption{Angular distributions at $\ln\Lambda=7$ and $N=500$, shown as the
  azimuth-integrated density $2\pi f(\chi)$ with $\omega$ defined by
  Eq.~\eqref{eq:moliere-width}. Yellow solid, red solid, and green dotted curves
  denote direct collision-by-collision Monte Carlo, the analytic Boltzmann
  result, and the Nanbu--P\'erez (NP) distribution, respectively.
  (a) Forward and intermediate angles, $0<\chi/\omega<3$, with the red curve
  from Eq.~\eqref{eq:laguerre-core}. (b) Rare-event range,
  $2<\chi/\omega<10$, with the red curve from
  Eq.~\eqref{eq:analytic-tail}. The black dashed curve is the normalized
  single-collision law,
  $2\pi N\sigma_{\rm bi}(\chi)/\sigma_t
  =N/[2\Lambda^2\sin^4(\chi/2)]$.}
  \label{fig:core-tail-emergence}
\end{figure}

For comparison, we use the widely employed Nanbu--P\'erez (NP) Monte Carlo
method, which reproduces Fokker--Planck dynamics in the small-angle limit through
moment matching~\cite{nanbu_theory_1997,bobylev_theory_2000,perez_improved_2012}.
Figure~\ref{fig:core-tail-emergence}
tests both angular limits against direct collision-by-collision Monte Carlo 
sampling and this Fokker--Planck benchmark for $\ln\Lambda=7$ and $N=500$.
In the forward sector [Fig.~\ref{fig:core-tail-emergence}(a)], the Laguerre
representation follows the Monte Carlo distribution from the peak through
$\chi/\omega=3$. The NP distribution, despite reproducing the
prescribed mean transport, gives a substantially lower forward peak and places
more probability at intermediate deflections, confirming the coefficient analysis
above at the distribution level.

Figure~\ref{fig:core-tail-emergence}(b) shows that the NP distribution decays
too rapidly to retain the rare-event population. The leading single-collision
tail underestimates the distribution by more than a factor of four near the
lower end of the panel and agrees only in the far tail. The second-order form
approaches the Monte Carlo result as the $\chi^{-4}$ behavior emerges and
continues to follow it at larger angles. Extending the elementary large-$k$
series through $k^{-8}$ increases the error in the overlap
$2<\chi/\omega<3$, making the second-order form the optimal explicit
truncation for asymptotic matching.
Within the logarithmic accuracy of the screened-Coulomb treatment, the core and
this optimally truncated tail can be combined through the intrinsic matching
developed below into a practical full-range representation. Further analytic
improvement would require a nonexplicit matching based on the full
special-function form of $H(x)$ in the Hankel integral, which is impractical for
repeated sampling in large-scale kinetic simulations (Appx. \ref{app:core-tail}).

Equations~\eqref{eq:laguerre-core} and~\eqref{eq:analytic-tail} describe two
limits of the same GS coefficient sequence. The core is controlled by the band
of Legendre orders over which weak deflections accumulate, whereas the tail is
fixed by the nonanalytic small-$x$ imprint of the binary singularity. The next
subsection determines where these limits overlap and how their domains change
with the collision parameters.

\subsection{Intrinsic crossover}
\label{sec:intrinsic-crossover}

The core and tail derived above overlap but neither is uniformly accurate over
the entire angular range. Their overlap must therefore be determined from the
two asymptotic branches. To make this construction explicit, remove
the common geometric factor and write
\begin{equation}
\begin{aligned}
  q_0(\chi)&=\frac{1}{2\pi\omega^2}
  \sum_{n=0}^{n_{\max}}(-1)^n h_n
  \ell_n\!\left(\frac{\chi^2}{\omega^2}\right),\\
  q_\infty(\chi)&=\frac{C_\infty}{\chi^4}
  \left[1+\frac{32N}{\Lambda^2\chi^2}
  \left(\ln\chi+\ln\Lambda'\right)\right].
  \label{eq:matching-branches}
\end{aligned}
\end{equation}
Here $q_0$ is the finite-order form of Eq.~\eqref{eq:laguerre-core}, while
$q_\infty$ retains the two tail orders in Eq.~\eqref{eq:analytic-tail}, and
$\ln\Lambda'=\ln\Lambda-\ln2-1/2$. A full-range approximation to the smooth
scattered population is then
\begin{equation}
  g_{\rm comp}(\chi)=C\sqrt{\frac{\chi}{\sin\chi}}
  \begin{cases}
    q_0(\chi),&0<\chi\leq\chi_c,\\
    q_\infty(\chi),&\chi_c<\chi<\pi,
  \end{cases}
  \label{eq:composite-distribution}
\end{equation}
where $C$ enforces
$\int g_{\rm comp}(\chi)\,\dd\Omega=1-e^{-N}$. The unscattered weight
$e^{-N}\delta_\Omega$ from Eq.~\eqref{eq:GS-smooth-decomposition} is retained
separately.

Continuity of both probability density and slope gives
\begin{equation}
  q_0(\chi_c)=q_\infty(\chi_c),\qquad
  q_0'(\chi_c)=q_\infty'(\chi_c).
  \label{eq:intrinsic-matching}
\end{equation}
The equality of logarithmic slopes first fixes $\chi_c$, and the equality of the
values then fixes $C_\infty$. Thus neither quantity is an empirical transition
parameter. More importantly, $\chi_c$ does not separate two collision
mechanisms. It marks the point of local overlap between two asymptotic representations
of the same GS distribution~\eqref{eq:GS-hankel-kernel}. A particle contributing beyond 
$\chi_c$ may have experienced a close deflection together with any number of weaker 
collisions, as already encoded by the second term of Eq.~\eqref{eq:analytic-tail}.

\begin{figure}[t]
  \centering
  \includegraphics[width=\columnwidth]{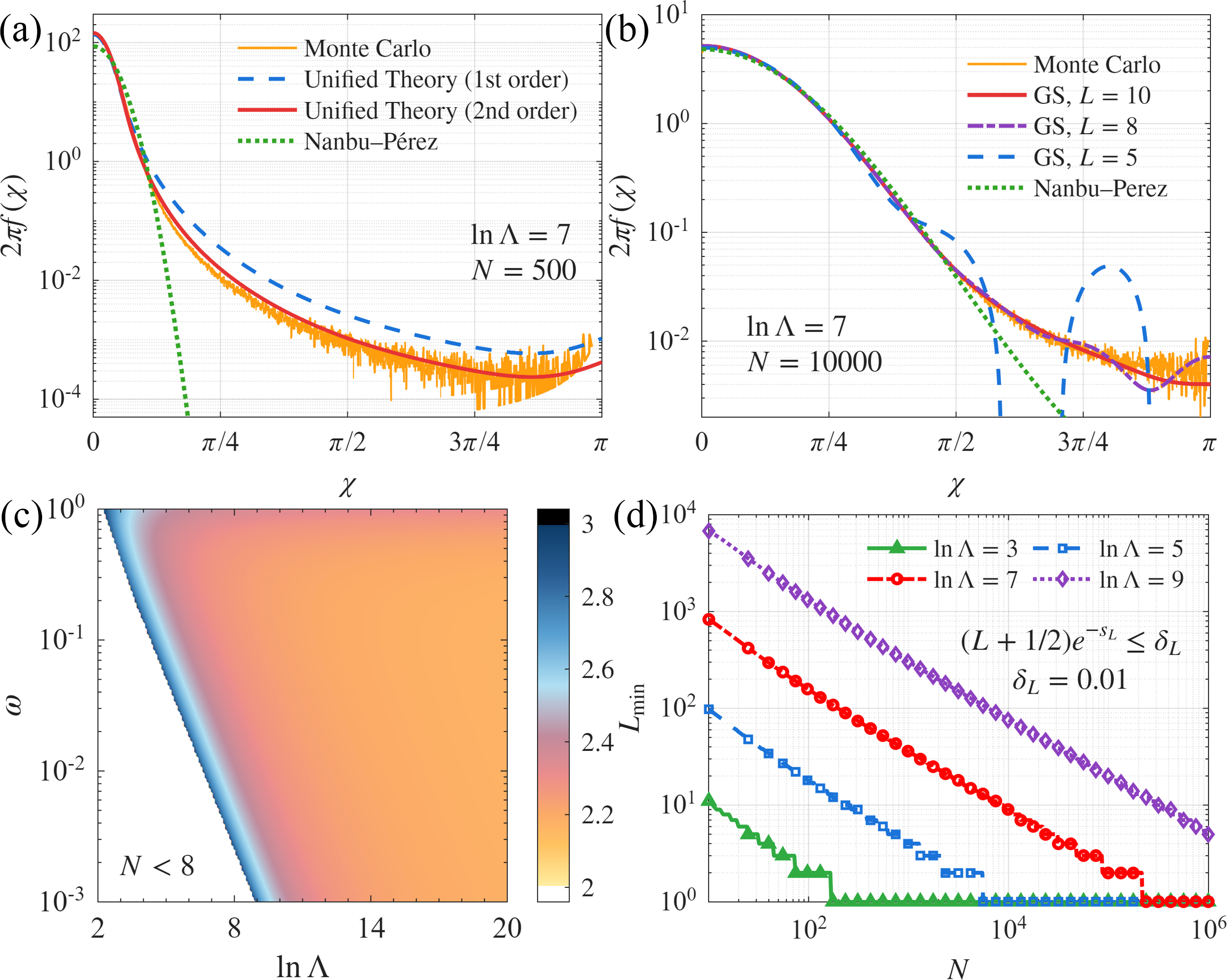}
  \caption{Intrinsic crossover and statistical regimes.
  (a) Full-range distributions at $\ln\Lambda=7$ and $N=500$ from direct
  Monte Carlo, the NP model, and the analytic Boltzmann
  representation retaining the first or first two tail orders.
  (b) Distributions at $\ln\Lambda=7$ and
  $N=10^4$ from Monte Carlo, NP, and the GS series truncated at
  $L=5$, $8$, and $10$. (c) Scaled crossover $\chi_c/\omega$ over the
  $(\ln\Lambda,\omega)$ plane. The blank region lies outside the \(N\gg1\) 
  range of the asymptotic construction.
  (d) Minimum positive integer $L$ satisfying
  Eq.~\eqref{eq:low-order-criterion} for $\delta_L=0.01$.}
  \label{fig:matching-regimes}
\end{figure}

Figure~\ref{fig:matching-regimes}(a) tests the resulting full-range
distribution. At $\ln\Lambda=7$ and $N=500$, retaining the second
tail order brings the composite distribution into close agreement with direct
collision-by-collision Monte Carlo over the entire angular range. Retaining
only the leading $\chi^{-4}$ term instead substantially overestimates the
rare-event probability. The NP distribution is included as a widely used
moment-matched small-angle benchmark. Its discrepancy is not confined to the
far tail, but extends through the forward and intermediate-angle sectors.

The matching conditions in Eq.~\eqref{eq:intrinsic-matching} yield a unique
crossover angle $\chi_c$ and a composite distribution that remains positive over
$\chi\in(0,\pi)$. Across the wide parameter range examined in 
Fig.~\ref{fig:matching-regimes}(c), this angle obeys
\begin{equation}
  2<{\chi_c}/{\omega}<3.
  \label{eq:crossover-band}
\end{equation}
Although $\omega$ is derived from the forward core, it also sets the angular 
scale at which rare events become dominant. The location~\eqref{eq:crossover-band}
explains why the second tail order is required: matching occurs before the leading 
$\chi^{-4}$ term has reached its far-tail limit. The second-order tail is the
optimal explicit asymptotic truncation in this region.
Appendix~\ref{app:core-tail} shows that this correction emerges from the same
expansion as the forward core. Omitting it changes low-order moments of the cumulative 
angular distribution, as demonstrated in Sec.~\ref{sec:alpha-validation}.

\subsection{Three statistical regimes within unified Boltzmann framework}
\label{sec:statistical-regimes}
The same analysis organizes the cumulative process into three statistical
regimes. For $N=O(1)$, the Poisson weights of zero, one, and a few collisions
remain individually visible. In this limit, the condition $N\gg1$ underlying
the asymptotic construction is not satisfied, and a smooth cumulative core has
not yet emerged, as shown in Appx.~\ref{app:core-tail}. The natural
description is therefore the finite sum of successive binary events. For
$1\ll N\ll\Lambda^2$ with $\omega\ll1$, many small-angle deflections form a
narrow core while the binary singularity remains visible through the $\chi^{-4}$
tail. This is the core--tail coexistence regime described by
Eq.~\eqref{eq:composite-distribution}.

At still larger $N$, the original GS series becomes simple for the opposite
reason. The small-$x$ form of the binary-collision coefficients gives
\begin{equation}
  s_l\simeq\frac{2N\ln\Lambda}{\Lambda^2}\rho^2
  \left[1-\frac{\ln\rho+O(1)}{\ln\Lambda}\right],
  \qquad \rho=l+\frac12.
  \label{eq:large-N-coefficients}
\end{equation}
Once $2N\ln\Lambda/\Lambda^2$ approaches unity, $e^{-s_l}$ decreases rapidly
with $l$ and only the first few Legendre orders contribute appreciably. The
distribution is then broad and progressively relaxed, with its high-order
structure strongly suppressed. We refer to this as the low-order relaxation
regime. Large-order Legendre-to-Bessel asymptotics are no longer
accurate~\cite{winterbon_finite_angle_1987}, whereas direct truncation of the
discrete GS series retains the relevant coefficients exactly.

These regimes are asymptotic domains rather than sharp phases. For a prescribed truncation 
tolerance $\delta_L$, the onset of the low-order regime can be stated as
\begin{equation}
  \left(L+1/2\right)e^{-s_L}\leq\delta_L,
  \label{eq:low-order-criterion}
\end{equation}
which provides an operational test of whether retaining terms through order $L$
is sufficient at that tolerance. Equation~\eqref{eq:low-order-criterion} does not
identify a universal point at which a Fokker--Planck approximation becomes ``valid''. Instead, it
defines an accuracy-dependent boundary for the low-order GS representation.

Panels (b) and (d) of Fig.~\ref{fig:matching-regimes} make this boundary
quantitative. At $\ln\Lambda=7$ and $N=10^4$, the $L=10$ GS truncation follows the 
Monte Carlo distribution, whereas $L=8$ is insufficient and the 
$L=5$ case even develops negative intervals, seen as gaps on the
logarithmic plot [Fig.~\ref{fig:matching-regimes}(b)]. The NP result now agrees
much more closely at moderate angles than in panel (a), as expected when the
distribution enters the low-order relaxation regime, but it still misses the
finite-angle probability. Figure~\ref{fig:matching-regimes}(d) generalizes this
test by plotting the minimum positive integer $L$ required for
$\delta_L=0.01$. The decrease with $N$ shows how the exact GS representation becomes
progressively shorter, while increasing $\ln\Lambda$ delays that reduction exponentially.

The most compact representation changes with collision number. The Poisson
expansion is short at $N=O(1)$, the composite form resolves the distribution at
intermediate $N$, and the exact GS series becomes short at large $N$. Each is a
representation of the same Boltzmann collision operator in a different
statistical limit.

\subsection{General elastic scattering and identical particles}
\label{sec:identical-particles}

The analysis above does not rely on the Rutherford formula itself. It requires
only an axisymmetric elastic differential cross section. Once that cross
section is supplied, Eqs.~\eqref{eq:general-asymptotic-coefficient}--
\eqref{eq:bounded-coefficient} translate its small-angle behavior into the GS
expansion coefficients and hence into the cumulative angular distribution. The
leading singularity determines the common structure. In particular,
charged-particle cross sections proportional to $\theta^{-4}$ at small angles
share the leading rescaling of the expansion coefficients and generate a
$\chi^{-4}$ rare-event tail. The remaining angular dependence enters through 
the subleading Laurent coefficients $c_{j>-p+1}$, whose contribution decays as $\mathcal{A}\rightarrow\infty$. 
The same framework therefore captures both the common consequences of the
leading small-angle singularity and the physics specific to each binary interaction.

Scattering between identical electrons provides a stringent example. Quantum
statistics enters electron collisions in two distinct ways: through Pauli
blocking of the final states and through exchange in the scattering amplitude.
In a nondegenerate plasma, $k_BT\gg E_F$, the blocking factors approach unity,
so the collision operator reduces to the Boltzmann form. Exchange nevertheless
remains in the physical differential cross section, which is given by
\Moller{} rather than classical Rutherford scattering~\cite{uehling_transport_1933}.
The original Nanbu-P\'erez construction treats the collision partners as
distinguishable~\cite{nanbu_theory_1997}. Its angular update can be applied to
electron macroparticles, but using a classical Coulomb cross section does not
incorporate the exchange physics of \Moller{} scattering~\cite{perez_improved_2012}.

In the center-of-momentum frame, the spin-averaged \Moller{} cross section can
be written as~\cite{raczka_moller_1958,sakurai_modern_2020}
\begin{equation}
  \frac{\dd\sigma_{\rm M}}{\dd\Omega}
  =\left(\frac{2\alpha\hbar c}{E_{\rm CM}}\right)^2
  \left[
    \frac{1}{\sin^4\theta}
    +\frac{\xi-3/4}{\sin^2\theta}
    +\frac{\xi}{4}
  \right],
  \label{eq:moller-cross-section}
\end{equation}
where $E_{\rm CM}=2m_ec^2\sqrt{1+(p_\star/m_ec)^2}$,
$p_\star$ is the electron momentum in that frame, and
$\xi=[(m_ec/p_\star)^2+2]^{-2}\in(0,1/4)$. Its leading pole remains
$p=4$, but the subleading coefficient in Eq.~\eqref{eq:laurent-kernel} becomes
$c_{-1}=\xi-1/6$, instead of the Rutherford value $1/12$. The two processes
therefore belong to the same $\theta^{-4}$ small-angle class without having the
same finite-angle statistics.

For an unordered pair of identical particles, the final states at $\theta$ and
$\pi-\theta$ differ only by exchanging the two outgoing particles. The
nonredundant angular domain consequently ends at $\theta_M=\pi/2$. To represent
the result on the full sphere, let $g_{\rm M}(\chi)$ denote the normalized
smooth distribution obtained by temporarily assigning a label to one outgoing
electron. Exchanging the labels maps $\chi$ to $\pi-\chi$ while leaving the
physical final state unchanged. The unlabeled distribution must therefore
average the two assignments. Because
$P_l[\cos(\pi-\chi)]=(-1)^lP_l(\cos\chi)$, this symmetrization preserves
normalization and cancels every odd Legendre component:
\begin{equation}
  \tilde{\sigma}^{\rm id}_{2l+1}=0,
  \qquad
  g_{\rm M}^{\rm id}(\chi)
  =\frac12\left[g_{\rm M}(\chi)+g_{\rm M}(\pi-\chi)\right].
  \label{eq:identical-symmetrization}
\end{equation}

It follows directly that
$g_{\rm M}^{\rm id}(\chi)=g_{\rm M}^{\rm id}(\pi-\chi)$ and
$\partial_\chi g_{\rm M}^{\rm id}(\pi/2)=0$. Thus exchange symmetry enters the
complete hierarchy of GS coefficients through the physical \Moller{} cross
section.

The coefficient rescaling, core--tail analysis, and intrinsic matching derived
for Rutherford scattering therefore carry over without changing their
structure. Only the collision-specific inputs in
Eq.~\eqref{eq:general-asymptotic-coefficient} are replaced. This extension is
restricted here to nondegenerate elastic scattering. Degenerate plasmas, 
for which Pauli blocking depends on the evolving occupation, require 
the full quantum collision operator and lie outside the present framework. The 
\Moller{} cross section~\eqref{eq:moller-cross-section} is used in
Sec.~\ref{sec:electron-transport} as the collision input for the fast-electron
transport example.

\section{Physical Consequences of Boltzmann Collision Statistics}
\label{sec:physical-consequences}

\subsection{Fast-electron transport}
\label{sec:electron-transport}

\begin{figure*}[!ht]
  \centering
  \includegraphics[width=0.9\textwidth]{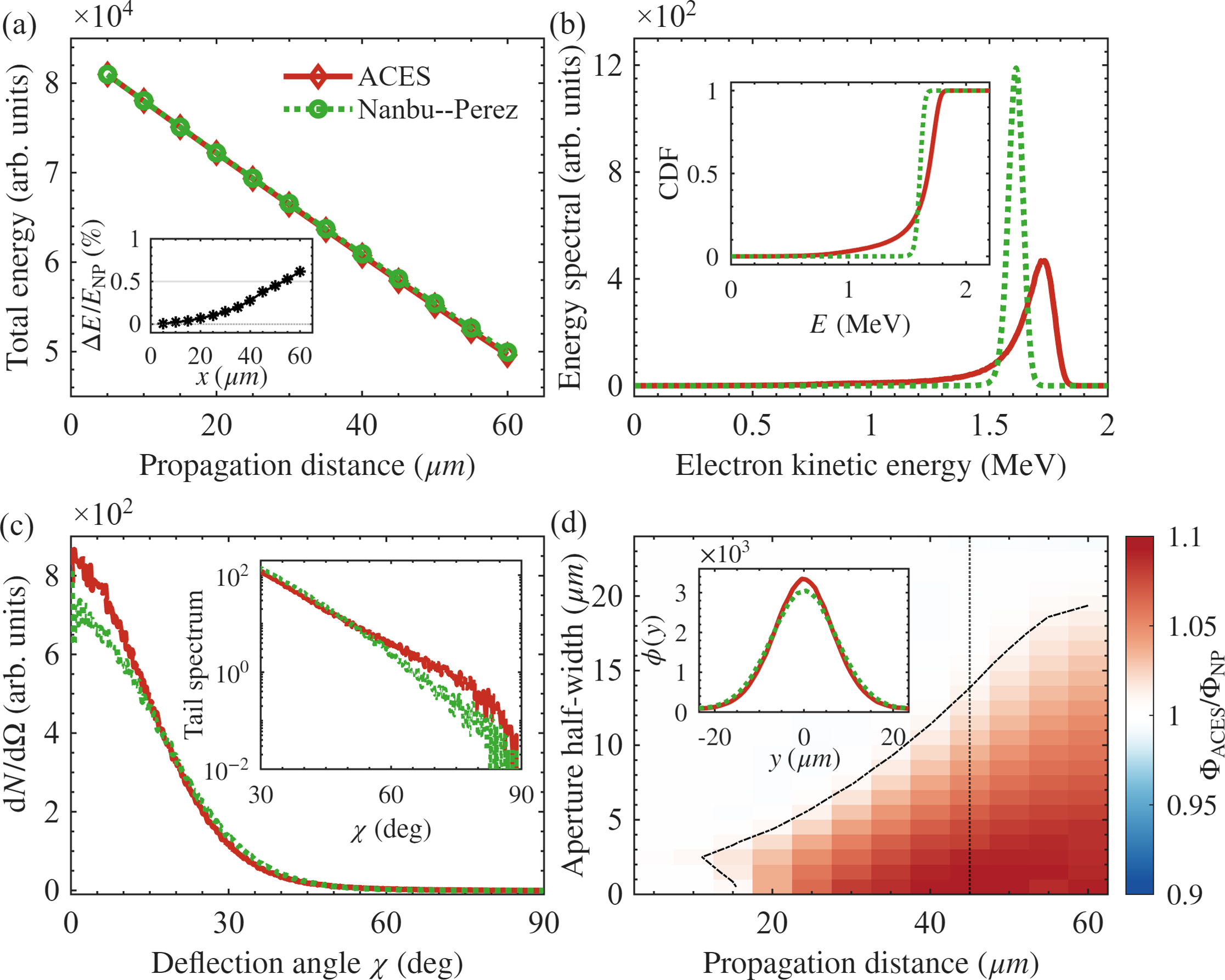}
  \caption{Boltzmann collision statistics in fast-electron transport.
  Red solid curves show the Boltzmann-based ACES treatment and green dotted
  curves the NP small-angle treatment.
  (a) Total kinetic energy transmitted through successive detector planes.
  The inset shows the relative difference between the two calculations.
  (b) Electron energy spectra at $x=45~\mu\mathrm{m}$, with the corresponding
  cumulative distribution functions (CDFs) in the inset.
  (c) Absolute angular spectra $\dd N/\dd\Omega$, at the same plane.
  The inset resolves the far-angle population on a logarithmic scale.
  (d) Ratio of the time-integrated kinetic-energy flux $\Phi(a,x)$
  transmitted through a centered aperture of half-width $a$. The vertical
  dotted line marks
  $x=45~\mu\mathrm{m}$, the black dashed contour marks $\Phi_{\rm ACES}/\Phi_{\rm NP}=1.01$,
  and the inset compares the transverse energy-flux $\phi(y)$ profiles at that plane.}
  \label{fig:electron-transport}
\end{figure*}

The analytic results above can be sampled through the any-angle cumulative
elastic scattering (ACES) collision treatment. The binary update is summarized
in Appx.~\ref{app:sampling}, and its implementation and sampling tests are given
in the SM, Secs. II and III~\cite{SM}. We use ACES to test whether
distributional differences remain observable when mean transport agrees. A
monoenergetic $2$-MeV beam of transverse half-width
$3~\mu\mathrm{m}$ along
$+x$ is continuously injected into uniform DT fuel at $T_e=T_i=1$~keV and
$\rho_{\rm DT}\simeq100~\mathrm{g\,cm^{-3}}$, a representative dense-plasma
transport problem relevant to fast ignition~\cite{tabak_ignition_1994,
tabak_fast_2006,robinson_theory_2014}. The calculation is performed with the
fully relativistic PIC code KLAPS~\cite{wang_integrated_2015}. Electron--electron
collisions in ACES use the spin-averaged \Moller{} cross
section introduced in Sec.~\ref{sec:identical-particles}, and other collision processes 
use the relativistic Coulomb-scattering cross sections derived by Frankel 
{\it et al.}~\cite{frankel_energy_1979}. The reference calculation uses the widely 
adopted NP small-angle treatment~\cite{nanbu_theory_1997,perez_improved_2012}. 
All other physical and numerical parameters are identical. We also repeated the ACES 
calculation with the corresponding classical Coulomb cross section for $e-e$ scattering. 
Its results are almost indistinguishable from the \Moller{}-based ACES curves in this example. 
The differences examined below thus arise from the angular statistics rather than from the 
exchange correction. The detectors record forward crossings at planes separated by
$5~\mu\mathrm{m}$. Simulation and diagnostic details are given in 
Appx.~\ref{app:simulation-details}.

We first isolate the information already fixed by low-order transport.
Figure~\ref{fig:electron-transport}(a) shows the total kinetic energy carried
across each detector plane, integrated over the full transverse domain. The
two curves remain nearly indistinguishable throughout $60~\mu\mathrm{m}$ of
propagation, with a relative difference below $0.7\%$. At the representative
plane $x=45~\mu\mathrm{m}$, their mean electron energies differ by only
$0.4\%$. Thus both descriptions give essentially the same mean collisional
energy loss, as expected from their common leading Coulomb transport rate.
Differences in the resolved distributions therefore cannot be attributed to a
different mean stopping rate.

The underlying collision statistics are nevertheless markedly different.
Figure~\ref{fig:electron-transport}(b) shows that NP retains a narrow Gaussian 
spectrum around the mean energy, whereas ACES produces a strongly
asymmetric distribution with a long low-energy wing, reflecting the broader finite-step
energy-transfer statistics induced by the Boltzmann angular distribution
~\cite{lee_quantum_2026}. The standard deviation
of the latter is approximately $6.2$ times larger at $45~\mu\mathrm{m}$ even
though the two means agree. This energy straggling is accompanied by a
full-range change in direction space [Fig.~\ref{fig:electron-transport}(c)].
ACES places about $18\%$ more particles within $\chi<5^\circ$, about $6\%$ 
fewer in the intermediate interval $\chi\in(15^\circ,35^\circ)$, and more than 
twice as many beyond $60^\circ$. The resulting distribution has a sharper
forward core, a depleted intermediate-angle population, and a sparse but
heavier far-angle tail, reproducing the coupled redistribution derived in
Sec.~\ref{sec:core-tail-emergence}.

The forward enhancement has a direct spatial consequence. For a centered
detector of half-width $a$, define the transmitted kinetic-energy flux
\begin{equation}
  \Phi(a,x)=\int\dd t\int_{-a}^{a}\dd y
  \int\dd E\,E\,\mathcal{J}(E,y;x,t),
  \label{eq:aperture-energy-flux}
\end{equation}
where $\mathcal{J}$ is the differential electron flux through the plane at
$x$, so that $\Phi$ is accumulated over the diagnostic interval.
Figure~\ref{fig:electron-transport}(d) maps
$\Phi_{\rm ACES}/\Phi_{\rm NP}$ as a function of propagation distance and
aperture. Integration over the full transverse domain returns the near-unity
ratio already established in panel (a), whereas a centered aperture of
half-width $3~\mu\mathrm{m}$ receives approximately $10\%$ more energy at
$x=45~\mu\mathrm{m}$. The transverse profiles in the inset identify the
origin of this increase: ACES produces a higher and narrower
central beam profile than the NP reference.

This behavior is counterintuitive if finite-angle events are regarded only as
an additional source of beam spreading. The Boltzmann--GS statistics retain a
heavier far-angle population, but they simultaneously transfer a
larger probability into the forward core at the expense of intermediate
deflections. For a finite acceptance, the latter redistribution dominates and
improves beam concentration while leaving the mean energy loss almost
unchanged. Low-order transport determines how much energy is lost on average,
while the scattering distribution determines where the surviving energy is
delivered.

\subsection{Alpha-particle knock-on tails}
\label{sec:alpha-validation}

Complementing the spatial-transport test above, we next examine how Boltzmann
collision statistics generate suprathermal ions in a homogeneous fusion
plasma. A close collision of a fusion-born alpha particle can transfer a
substantial fraction of its energy to a single D or T ion, producing the
knock-on tails observed in burning-plasma
diagnostics~\cite{fisher_fast_1994,kallne_observation_2000,
hartouni_evidence_2023,gatu_johnson_learning_2024}. The macroscopic
interpretation of these suprathermal populations remains unsettled. One recent
kinetic study attributed non-Maxwellian ion signatures and earlier ignition to
large-angle alpha--DT collisions~\cite{xue_mechanisms_2025}, whereas a
full-capsule calculation reproduced the alpha-knock-on-neutron signal but found
that these collisions could not explain the anomalous neutron spectral
shift~\cite{van_de_wetering_particle_2025}. Accurately resolving the knock-on
tail is therefore a prerequisite for assessing its contested role in burn
dynamics and neutron diagnostics.

We initialize a uniform DT plasma at
$\rho_{\rm DT}\simeq50~\mathrm{g\,cm^{-3}}$ and
$T_e=T_i=5$~keV with an isotropic population of $3.54$-MeV alpha particles,
$n_\alpha=0.05n_{\rm DT}$. The Maxwell solver is disabled. The system is evolved
in a two-dimensional periodic PIC domain. The ACES calculation uses the distribution
\eqref{eq:composite-distribution}, including the second-order tail in
Eq.~\eqref{eq:analytic-tail}, and the reference calculation again uses the NP
small-angle treatment. All plasma and numerical parameters are otherwise
identical, with further details given in Appx.~\ref{app:simulation-details}.

\begin{figure}[t]
  \centering
  \includegraphics[width=\linewidth]{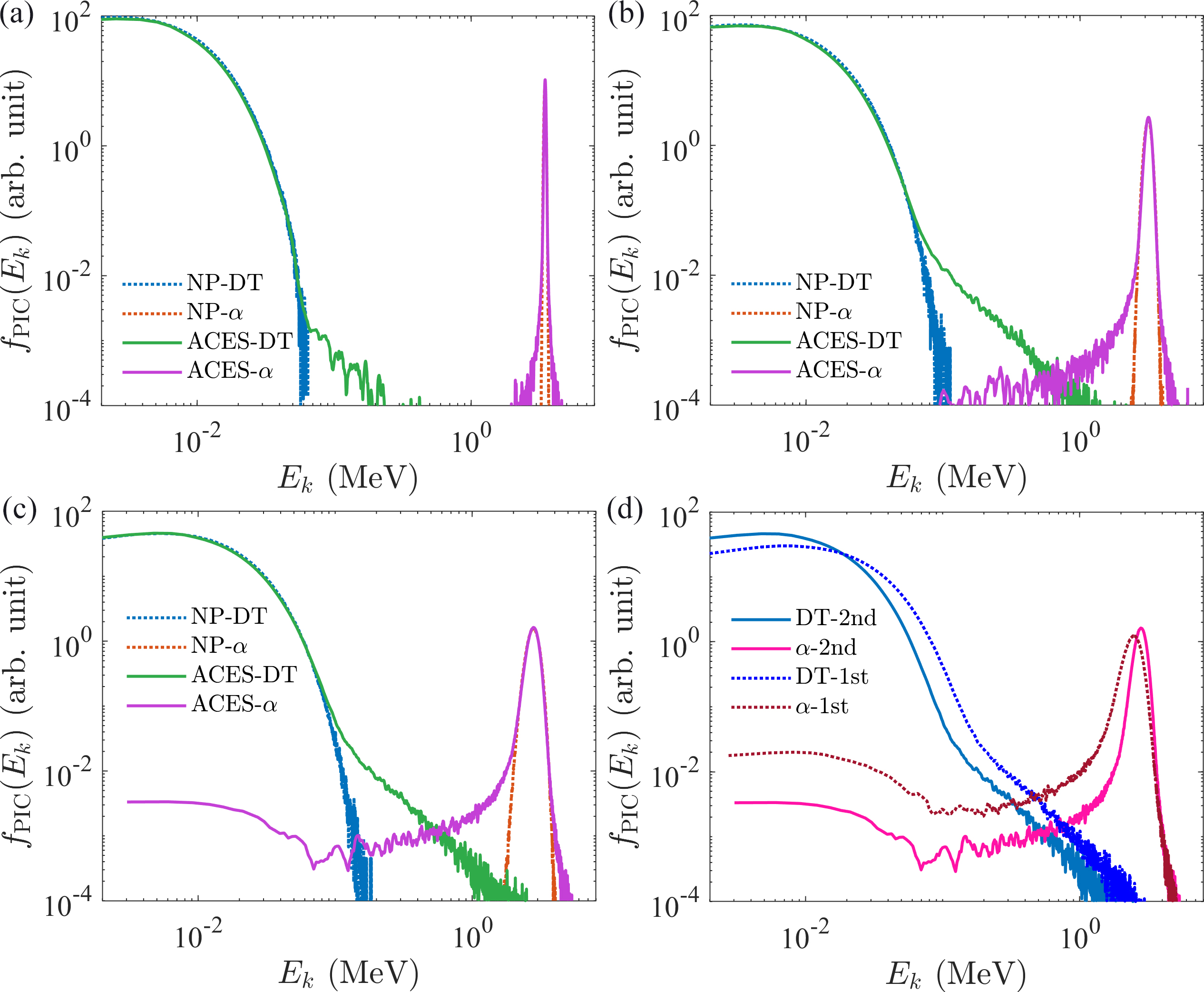}
  \caption{Energy response to alpha-particle scattering in uniform DT.
  (a)--(c) Energy distributions of DT ions and alpha particles at
  $t=0.05$, $1$, and $3$~ps. Solid curves show ACES with the 2nd-order
  tail retained, and dotted curves show NP. (d) ACES distributions at
  $t=3$~ps obtained with the second-order tail (solid) and the leading
  $\chi^{-4}$ term alone (dashed).}
  \label{fig:alpha-particle}
\end{figure}

The contrast appears first in the small population carrying a large amount of
energy per particle. By $t=0.05$~ps, ACES already produces DT ions beyond the
thermal distribution [Fig.~\ref{fig:alpha-particle}(a)]. As collisions
accumulate, this population develops into a broad tail extending through
$10^2$--$10^3$~keV at $t=3$~ps
[Figs.~\ref{fig:alpha-particle}(b) and~\ref{fig:alpha-particle}(c)]. The
corresponding alpha-particle distribution broadens toward lower energies,
recording the complementary large transfers. NP reproduces the thermal DT
population but generates neither this knock-on tail nor the accompanying
alpha-particle broadening. The simultaneous agreement of the thermal
population and disagreement of the knock-on tail show that the GS coefficient
hierarchy retains low-order relaxation and rare energy-transfer statistics
without an additional moment-matching equation.

The same benchmark further tests whether the finite-order angular
representation preserves this consistency. Figure~\ref{fig:alpha-particle}(d)
compares ACES calculations retaining either one or two tail terms in
Eq.~\eqref{eq:analytic-tail}. The leading $\chi^{-4}$ term alone overestimates 
the probability of finite-angle scattering events in the crossover region identified 
in Sec.~\ref{sec:intrinsic-crossover}. Accumulated over many alpha--ion encounters, 
that error produces excessive alpha energy deposition and shifts even the bulk DT 
distribution. Retaining the second-order term removes this spurious bulk change
while preserving the physical knock-on tail. This observable test confirms the
optimal second-order truncation found from the angular analysis. Accurate
crossover probabilities are required not only for the far tail but also to
prevent their accumulated error from shifting the dominant population.

\section{Discussion}
\label{sec:discussion}

The results distinguish three levels of fidelity in collision modeling.
Pairwise conservation guarantees neither the correct mean exchange nor the
correct collision statistics. Matching relevant transport moments, such as
$\langle1-\cos\chi\rangle$, restores low-order transport but leaves the higher
GS coefficients unconstrained. Distributional fidelity requires all
coefficients $\langle P_l(\cos\chi)\rangle=e^{-s_l}$ that remain appreciable
during the collision step. The two applications expose this distinction:
nearly identical mean energy loss can
coexist with different directional spectra and local flux, while agreement in
the thermal population can coexist with a missing knock-on tail. A description
adequate for mean relaxation therefore need not be adequate for finite
acceptance, threshold crossing, or suprathermal yield.

The NP treatment shows directly why moment-based models can be highly
successful without being distributionally equivalent to Boltzmann scattering.
Its finite-step distribution is
~\cite{nanbu_theory_1997,perez_improved_2012}
\begin{equation}
  f_{\rm NP}(\chi)=\frac{A}{4\pi\sinh A}e^{A\cos\chi},
  \quad \coth A-\frac{1}{A}=e^{-s_1}.
  \label{eq:nanbu-distribution}
\end{equation}
This is the von Mises--Fisher distribution, the maximum-entropy distribution on
the sphere when only the first directional moment is prescribed
\cite{mardia_directional_1999}. Expanding Eq.~\eqref{eq:nanbu-distribution} in
Legendre polynomials gives
\begin{equation}
  f_{\rm NP}(\chi)=\sum_{l=0}^{\infty}\frac{2l+1}{4\pi}
  a_l^{\rm NP}P_l(\cos\chi),
  ~a_l^{\rm NP}=\frac{A i_l(A)}{\sinh A},
  \label{eq:nanbu-legendre}
\end{equation}
where $i_l$ is the modified spherical Bessel function. The corresponding GS
expansion coefficient is $a_l^{\rm GS}=e^{-s_l}$. Their ratio is therefore
\begin{equation}
  \mathcal R_l\equiv\frac{a_l^{\rm NP}}{a_l^{\rm GS}}
  =\frac{A i_l(A)}{\sinh A}\,e^{s_l}.
  \label{eq:coefficient-ratio}
\end{equation}

\begin{figure}[t]
  \centering
  \includegraphics[width=\columnwidth]{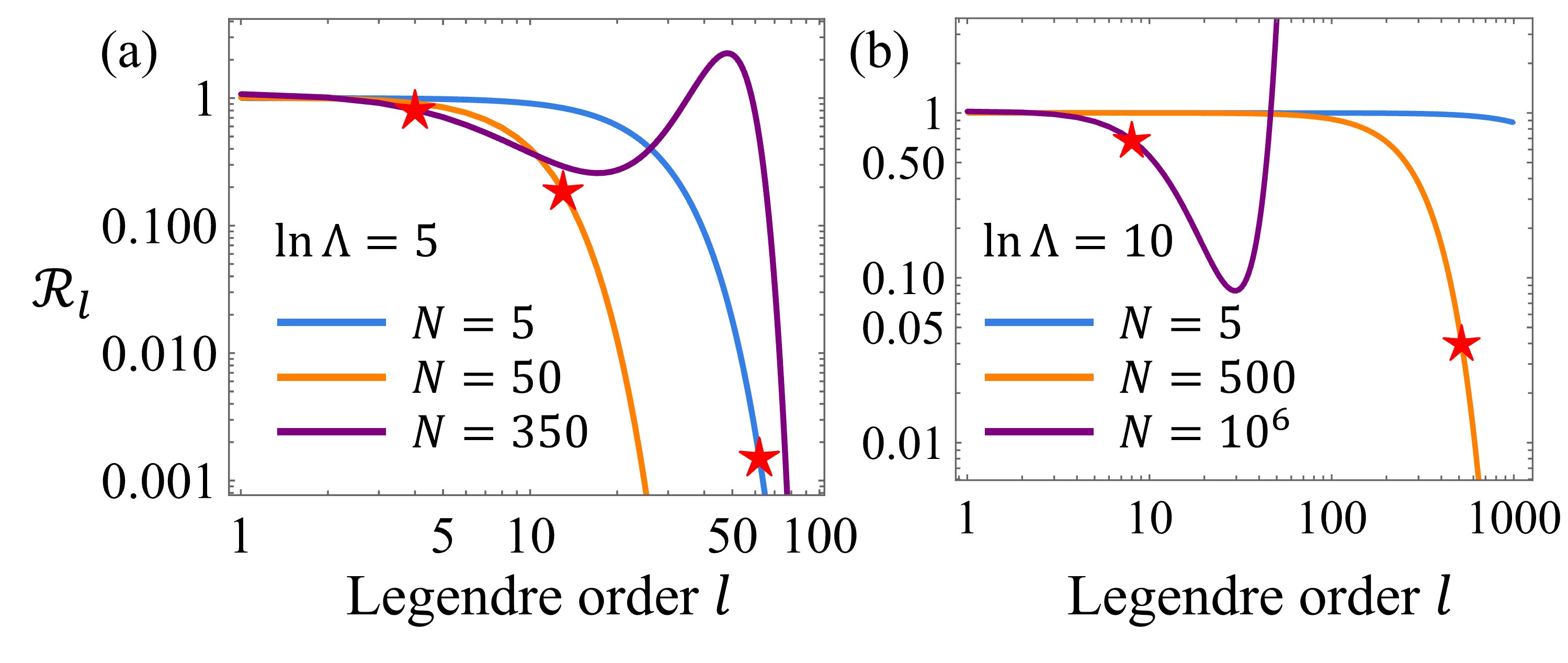}
  \caption{Coefficient-level comparison of the Nanbu--P\'erez and GS
  distributions for (a) $\ln\Lambda=5$ and (b) $\ln\Lambda=10$.
  The ratio $\mathcal R_l=a_l^{\rm NP}/a_l^{\rm GS}$ is shown for the indicated
  collision numbers. Each red star marks the first order satisfying
  $a_l^{\rm GS}=e^{-s_l}<0.1$ and is used only to indicate where the GS
  coefficient has become strongly attenuated. No star appears on the blue
  curve in panel (b) because this condition is not reached below $l=10^3$.
  Equations~\eqref{eq:rescaled-variable} and~\eqref{eq:rescaled-core-branch}
  predict it near $l=9.2\times10^3$.}
  \label{fig:coefficient-ratio}
\end{figure}

By construction, $\mathcal R_1=1$, so the two distributions predict the same
mean momentum transfer. The remaining NP coefficients, however, follow from
the assumed von Mises--Fisher shape rather than from the binary cross section.
The coefficient ratios in Fig.~\ref{fig:coefficient-ratio} remain remarkably
close to unity over a broad range of low orders, explaining why moment-based
descriptions can pass stringent transport tests. At higher orders they depart
from unity before the corresponding GS coefficients become strongly
attenuated, and the red stars mark where $a_l^{\rm GS}$ first falls below $0.1$.
Because the forward and
intermediate-angle profiles are reconstructed from the entire active
coefficient band rather than its first few members, agreement of one or even
several low-order moments does not establish distributional equivalence. The
differences in Figs.~\ref{fig:core-tail-emergence} and
\ref{fig:matching-regimes} are the angle-space manifestation of this unresolved
higher-order information.

At larger $N$, the higher GS coefficients are suppressed at progressively
lower $l$, leaving the shared low-order information to control an increasing
fraction of the distribution. This explains why NP approaches the Boltzmann
result in the low-order relaxation regime and why it can reliably reproduce mean
collisional transport more broadly.

The same distinction clarifies the relation of the present theory to methods
that combine soft and hard collisions. Conventional splicing prescriptions
append a single-collision tail to an assumed Fokker--Planck core
~\cite{turrell_self-consistent_2015,higginson_full-angle_2017}. Such a
construction restores finite-angle events but leaves the forward population
specified independently. In runaway-electron kinetics, relativistic
large-transfer Boltzmann operators have likewise been combined with a
Fokker--Planck treatment of small transfers~\cite{embreus_relativistic_2018}.
More recently, GCM has coupled cumulative and single scattering under an
explicit transport-moment constraint, thereby preserving mean transfer while
retaining a hard-event sector~\cite{ray_angus_monte-carlo_2026}. These
approaches are well suited to observables controlled by the moments and event
rates they enforce. The present theory instead determines the cumulative
distribution itself, which neither a prescribed transition nor a finite set of
moment constraints uniquely fixes. In particular, the sharper core observed in
Fig.~\ref{fig:electron-transport} cannot be inferred merely from adding a hard
tail. Whether a moment-constrained hybrid reproduces it must be established by
comparison with the full coefficient hierarchy rather than assumed from
transport agreement.

The construction is not specific to Coulomb scattering. For an axisymmetric
elastic cross section with $\sigma_{\rm bi}(\theta)\sim\theta^{-p}$, the leading
pole fixes the asymptotic rescaling of the entire GS coefficient hierarchy in
the small-cutoff limit, while the contribution from subleading Laurent terms vanish. The
corresponding hypergeometric-integral construction for different values of $p$
is given in Appx.~\ref{app:spectral-expansion}, and Coulomb scattering is the $p=4$ realization
developed here. The core, crossover, and rare-event statistics obtained above
are therefore consequences of a broader class of singular elastic-scattering
processes rather than features tied to a particular Coulomb model.

The present scope is the cumulative angular statistics generated by a
prescribed elastic cross section during one collision step, and evolution of the
relative energy and screening lies beyond it. The Fokker--Planck description is
an observable-dependent coarse graining of Boltzmann collision statistics, not
a specification of their full finite-step distribution. Equal mean transfer can conceal
different trajectories, spatial deposition, and threshold-crossing
probabilities because these observables probe different parts of the GS
coefficient hierarchy.

\section{Conclusion}
\label{sec:conclusion}

Starting from the GS series for a general axisymmetric elastic cross section
with a power-law small-angle singularity, we derived an asymptotic rescaling
that makes its coefficient hierarchy analytically accessible. For Coulomb
scattering, the rescaled coefficients yield the forward core, the
$\chi^{-4}$ rare-event tail, and their intrinsic crossover directly from the
binary cross section, without empirical transition parameters. The same
construction identifies single-collision, core--tail coexistence, and
low-order relaxation as controlled regimes of cumulative Boltzmann
scattering.

The results also establish the limit of transport-moment
equivalence. Matching the first transport moment can reproduce mean momentum
and energy transfer while leaving the cumulative distribution undetermined.
In fast-electron transport, this distinction appears as nearly identical mean
energy loss but a sharper forward population, enhanced beam concentration, and
modified local energy flux. In alpha-particle relaxation, the GS statistics
retain the thermal population while generating
the suprathermal knock-on tail. In both cases, low-order agreement leaves
spatial or rare-event observables undetermined.

The analysis extends to axisymmetric elastic cross sections with a small-angle
singularity, including identical-electron scattering described by the physical
\Moller{} cross section. ACES makes the resulting statistics available in
large-scale kinetic simulations. Relevant settings include the collimation and
deposition of fast electrons and alpha particles in inertial-confinement fusion
and fast ignition
\cite{robinson_theory_2014,zhang_double-cone_2020,du_modeling_2024,kallne_observation_2000},
tokamak runaway-electron dynamics
\cite{breizman_physics_2019,embreus_relativistic_2018}, identical-particle
scattering in cosmological plasmas
\cite{gould_kinetic_1981,dermer_binary_1985,arido_simulating_2025}, and
general-purpose Monte Carlo transport
\cite{kadri_incorporation_2009,kawrakow_accurate_2000,novak_new_2024}. More
broadly, the theory provides a direct basis for deciding when Fokker--Planck-level
information is sufficient and when trajectories, spatial deposition, or
rare-event yields require the underlying Boltzmann collision distribution.

\begin{acknowledgments}
The authors thank Dr.\ Rong-Zhen Guo and Dr.\ Tian-Xing Hu for constructive discussions.
This work was supported by the Shanghai Municipal Science and Technology Major Project, 
the Strategic Priority Research Program of Chinese Academy of Sciences 
(Grant No. XDA25050300, XDA25010100, XDA25010300), the National Natural Science Foundation of China
(Grant Nos. 12595363 and U24A2016) and the Fundamental Research Funds 
for the Central Universities, the Research Funds of Shanghai Jiao Tong University (WH220207203).
\end{acknowledgments}

\appendix

\section{Asymptotic Expansion of the Legendre Coefficients}
\label{app:spectral-expansion}

This appendix gives the hypergeometric representation and asymptotic
properties used in Sec.~\ref{sec:rescaled-spectrum}.  We retain the notation of
the main text: \(\rho=l+1/2\), \(\vartheta=\rho\theta/2\),
\(\vartheta_M=\rho\theta_M/2\), and
\(\lambda=\theta_M/\theta_m\).

\subsection{Hypergeometric representation}
\label{app:hypergeometric-representation}

The strong small-angle singularity is naturally represented by the Laurent
expansion in Eq.~\eqref{eq:laurent-kernel}, rather than by a Taylor series.
For a general pole order $p$, the asymptotic coefficient takes the form of
Eq.~\eqref{eq:general-asymptotic-coefficient}:
\begin{equation}
 \tilde{\sigma}_l\simeq
  \frac{1}{\mathcal A}\sum_{j=0}^{\infty}c_{-p+1+j}
  \int_{\theta_m}^{\theta_M}J_0(\rho\theta)
  \theta^{-p+1+j}\,\dd\theta+R_l,
 \label{eq:app-bessel-integrals}
\end{equation}
where \(c_{-p+1}=1\), \(\rho=l+1/2\), and \(R_l\) denotes the standard
remainder of the Legendre--Bessel asymptotic
reduction~\cite{szego_orthogonal_1975}. The lower limit \(\theta_m\) is fixed
by the relevant small-angle regularization.

For integer $p\geq2$, which includes the Coulomb class, the individual
Bessel integrals admit explicit forms in standard special functions. With
\(\vartheta=\rho\theta/2\), \(\theta^{-2k}~(k\geq1)\) terms satisfy
\begin{equation}
 \int \frac{J_0(\rho\theta)}{\theta^{2k}}\,\dd\theta
 =-\frac{{}_1F_2\left(\frac{1}{2}-k;1,\frac{3}{2}-k;-\vartheta^2\right)}{(2k-1)\theta^{2k-1}},
\end{equation}
whereas \(\theta^{-(2k+1)}~(k\geq0)\) terms satisfy
\begin{equation}
 \int \frac{J_0(\rho\theta)}{\theta^{2k+1}}\,\dd\theta
 =-\frac{1}{2\theta^{2k}}G_{1,3}^{2,0}\left(\vartheta^2\Bigg|
 \begin{matrix}
 k+1\\ 0,k;0
 \end{matrix}\right).
\end{equation}

The latter Meijer-$G$ function can be decomposed into a term proportional to
\(\vartheta^{2k}\ln\vartheta\), an entire generalized hypergeometric
function, and a polynomial of degree \(k-1\). 

We now specialize to the Coulomb class \(p=4\), including Rutherford and
\Moller{} scattering. For both kernels, the even-indexed Laurent coefficients
vanish. For Rutherford scattering,
\(\ln\Lambda=\ln\cot(\theta_m/2)\). The upper limit is
\(\theta_M=\pi\) for distinguishable particles and \(\theta_M=\pi/2\) for
identical particles. The first collision-specific Laurent coefficients are

{\small
\setlength{\tabcolsep}{3pt}
\centering
\begin{tabular}{c|cccc}
 & \(c_{-3}\) & \(c_{-1}\) & \(c_1\) & \(c_3\) \\ \hline
 Rutherford & \(1\) & \(1/12\) & \(1/480\) & \(-1/5760\) \\
 \Moller & \(1\) & \(\xi-1/6\) & \(\xi/2+1/480\) &
 \((23+27\xi)/1440\)
\end{tabular}\par
}
Their normalization factors are
\begin{align}
 \mathcal A_R&=\frac18\cot^2\frac{\theta_m}{2}
 =\frac{\Lambda^2}{8},
 \label{eq:app-AR}\\
 \mathcal A_M&=\frac{\cos\theta_m}{2\sin^2\theta_m}
 -\left(\frac14-\xi\right)\ln\cot\frac{\theta_m}{2}
 +\frac{\xi}{4}\cos\theta_m.
 \label{eq:app-AM}
\end{align}
Both reduce to \(\Lambda^2/8\) for \(\theta_m\ll1\). The required indefinite 
integrals for $p=4$ can be expressed as
\begin{align}
 &\int\frac{J_0(\rho\theta)}{\theta^3}\,\dd\theta
 =\frac{\rho^2}{4}\left[
 -\frac{1}{2\vartheta^2}-\ln\vartheta
 +\frac{\vartheta^2}{8}\mathcal F_{-3}(\vartheta^2)\right],
 \notag\\
 &\int\frac{J_0(\rho\theta)}{\theta}\,\dd\theta
  =\ln\vartheta-\frac{\vartheta^2}{2}
  \mathcal F_{-1}(\vartheta^2),\label{eq:app-indefinite-integrals}\\
 &\int\theta^{2k+1}J_0(\rho\theta)\,\dd\theta
  =\frac{\theta^{2k+2}}{2(k+1)}
  \mathcal F_{2k+1}(\vartheta^2),\quad k\geq0.\notag
\end{align}

Here, 
\begin{align}
 \mathcal F_{-3}(z)&={}_2F_3(1,1;2,3,3;-z),\notag\\
 \mathcal F_{-1}(z)&={}_2F_3(1,1;2,2,2;-z),
 \label{eq:app-hypergeometric-functions}\\
 \mathcal F_{2k+1}(z)&={}_1F_2(k+1;1,k+2;-z).\notag
\end{align}

These functions are entire in \(z\) and satisfy \(\mathcal F_j(0)=1\).
The additive constants in Eq.~\eqref{eq:app-indefinite-integrals} cancel
between the upper and lower limits. Substituting the screening cutoff 
\(\theta_m\) and the maximum scattering
angle \(\theta_M\) into Eq.~\eqref{eq:app-indefinite-integrals} gives
Eq.~\eqref{eq:app-bounded-integrals}.

\begin{align}
 &\mathcal I_{-3}(\lambda,\vartheta_M)
 =1-\frac{1}{\lambda^2}
 +\frac{\vartheta_M^2}{4\lambda^2}\label{eq:app-bounded-integrals}\\
 &\qquad\qquad\times\left[
 \vartheta_M^2\mathcal F_{-3}(\vartheta_M^2)
 -\frac{\vartheta_M^2}{\lambda^2}
 \mathcal F_{-3}\!\left(\frac{\vartheta_M^2}{\lambda^2}\right)
 -8\ln\lambda
 \right],\notag\\
 &\mathcal I_{-1}(\lambda,\vartheta_M)
 =\vartheta_M^2\mathcal F_{-1}(\vartheta_M^2)
 -\frac{\vartheta_M^2}{\lambda^2}
 \mathcal F_{-1}\!\left(\frac{\vartheta_M^2}{\lambda^2}\right)
 -2\ln\lambda,\notag\\
 &\mathcal I_{2k+1}(\lambda,\vartheta_M)
 =\frac{1}{k+1}\left[
 \mathcal F_{2k+1}(\vartheta_M^2)
 -\frac{\mathcal F_{2k+1}(\vartheta_M^2/\lambda^2)}
 {\lambda^{2k+2}}
 \right]\notag.
\end{align}

The functions \(\mathcal I_j\), defined by
definite integration, collect the dependence on these limits into
\(\lambda=\theta_M/\theta_m\) and \(\vartheta_M=\rho\theta_M/2\).

\subsection{Asymptotic properties and rescaling}
\label{app:function-properties}

The large-\(\vartheta\) properties of the hypergeometric functions are
\begin{equation}
 \vartheta^2\mathcal F_{-3}(\vartheta^2)
 \sim8(\ln\vartheta+\gamma_E-1)
 +\frac{4}{\vartheta^2}+O(\vartheta^{-7/2}),
 \label{eq:app-Fm3}
\end{equation}
\begin{equation}
 \vartheta^2\mathcal F_{-1}(\vartheta^2)
 \sim2(\ln\vartheta+\gamma_E)+O(\vartheta^{-3/2}),
 \label{eq:app-Fm1}
\end{equation}
\begin{equation}
 \mathcal F_{2k+1}(\vartheta^2)
 \sim\frac{\sin(2\vartheta-\pi/4)}
 {\sqrt{\pi}(k+1)\vartheta^{3/2}}
 +O(\vartheta^{-5/2}),~k\geq0.
 \label{eq:app-Fpositive}
\end{equation}
For \(\vartheta>2\), comparison with the exact hypergeometric functions gives
relative errors below \(2\%\) for the two singular contributions in
Eqs.~\eqref{eq:app-Fm3} and~\eqref{eq:app-Fm1}. Independently, the exact
combinations in Eq.~\eqref{eq:app-bounded-integrals} have the regular
zero-argument limits
\(\mathcal I_{-3}(\lambda,0)=1-\lambda^{-2}\),
\(\mathcal I_{-1}(\lambda,0)=-2\ln\lambda\), and
\(\mathcal I_{2k+1}(\lambda,0)=(1-\lambda^{-2k-2})/(k+1)\).

For \(\lambda\gg1\), the logarithmic and constant parts of the rapidly and
slowly varying terms in Eq.~\eqref{eq:app-bounded-integrals} cancel.  The
leading function consequently depends on \(l\) and \(\Lambda\) through
\[
 x=\frac{\vartheta_M}{\lambda}
 =\frac{\rho\theta_m}{2}\simeq\frac{\rho}{\Lambda}.
\]
At large \(x\), the corresponding functions behave as
\begin{align}
 \mathcal I_{-3}(\lambda,\lambda x)
 &\simeq\frac{\cos(2x+\pi/4)}{\sqrt{\pi}x^{3/2}}
 +O(x^{-5/2}),\label{eq:app-Im3-large}\\
 \mathcal I_{-1}(\lambda,\lambda x)
 &\simeq-\frac{\cos(2x+\pi/4)}{\sqrt{\pi}x^{3/2}}
 +O(x^{-5/2}),\label{eq:app-Im1-large}\\
 \mathcal I_{2k+1}(\lambda,\lambda x)
 &\simeq-\frac{\cos(2\lambda x+\pi/4)}
 {\sqrt{\pi}\lambda^{3/2}x^{3/2}}
 +O(x^{-5/2}).\label{eq:app-Ipositive-large}
\end{align}
Since \(\mathcal A\simeq\Lambda^2/8\), the
\(\mathcal I_{-3}\) term in Eq.~\eqref{eq:bounded-coefficient} determines the
large-\(\Lambda\) result.  Thus
\begin{equation}
 \tilde{\sigma}_l
 \simeq\tilde{\varsigma}(x).
 \label{eq:app-rescaled-coefficient}
\end{equation}

Taking the \(\lambda\to\infty\) limit of
\(\mathcal I_{-3}(\lambda,\lambda x)\) gives the explicit rescaled function
\begin{equation}
 \tilde{\varsigma}(x)
 \simeq 1+2x^2(\ln x+\gamma_E-1)
 -\frac{x^4}{4}\mathcal F_{-3}(x^2).
 \label{eq:app-rescaled-function}
\end{equation}
Its small-\(x\) expansion is
\begin{equation}
 \tilde{\varsigma}(x)
 \simeq 1+2x^2\ln x-2(1-\gamma_E)x^2-\frac{x^4}{4}
 +O(x^6),
 \label{eq:app-small-x}
\end{equation}
while Eq.~\eqref{eq:app-Im3-large} gives
\begin{equation}
 \tilde{\varsigma}(x)
 =\frac{\cos(2x+\pi/4)}{\sqrt{\pi}x^{3/2}}
 +O(x^{-5/2})
 \qquad (x\gg1).
 \label{eq:app-large-x}
\end{equation}
Equations~\eqref{eq:app-rescaled-coefficient}--\eqref{eq:app-large-x} are the
results used in Eqs.~\eqref{eq:rescaled-spectrum}--\eqref{eq:rescaled-oscillatory}.
Their agreement with the defining Legendre integral is shown in
Fig.~\ref{fig:rescaled-spectrum}.

The \(p=4\) Coulomb class is developed explicitly because it underlies the
subsequent analysis.

\section{Forward Core and Uniform Core--Tail Expansion}
\label{app:core-tail}

\subsection{Development of the forward core}

\begin{figure}[h]
 \centering
 \includegraphics[width=\columnwidth]{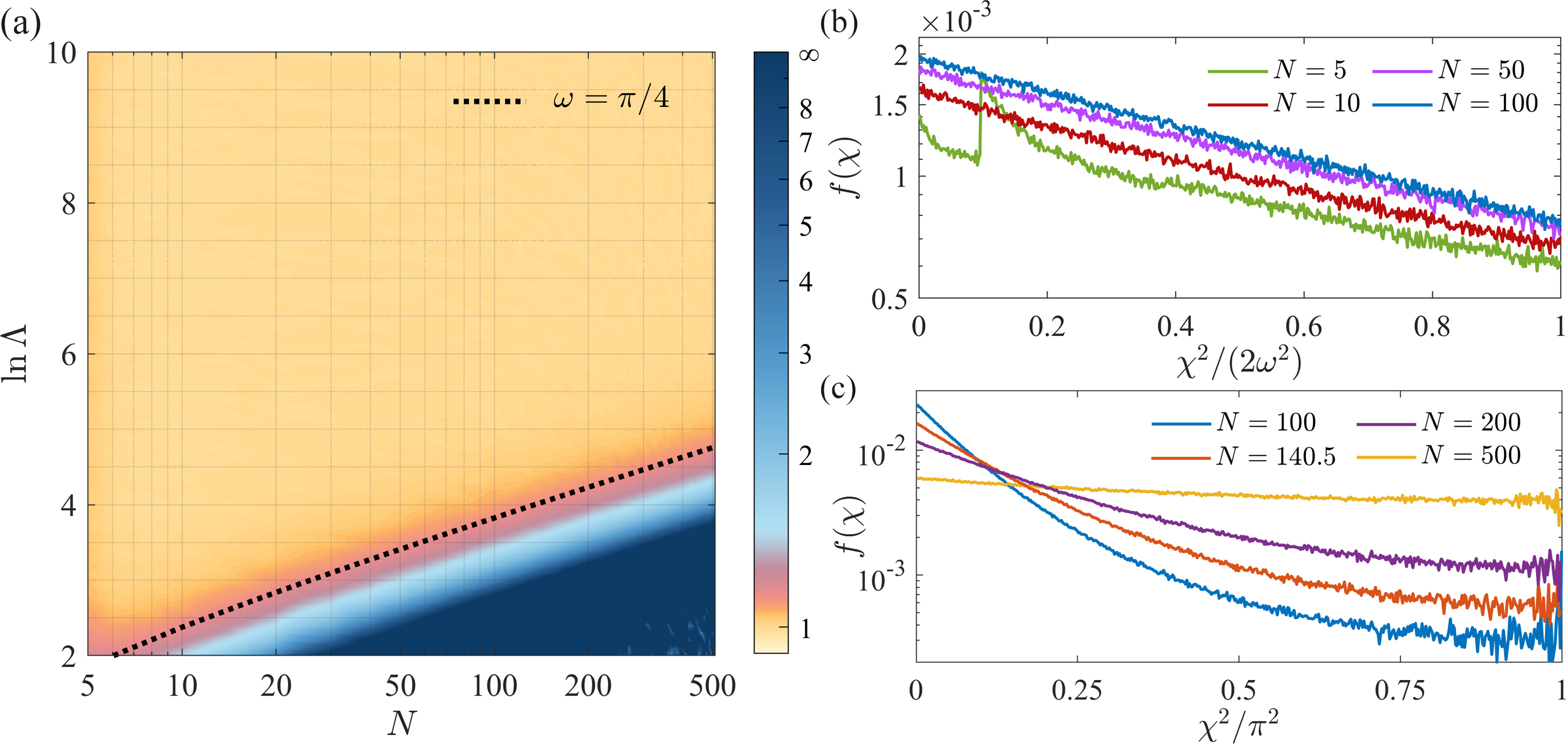}
 \caption{Moli\`ere width and the emergence of the forward core.
 (a) Ratio \(\omega_{\rm MC}^2/\omega^2\) obtained by fitting direct
 collision-by-collision Monte Carlo distributions. The dotted line marks
 \(\omega=\pi/4\).  (b) Forward distributions at \(\ln\Lambda=4\) for
 \(N=5,10,50,100\), shown against \(\chi^2/(2\omega^2)\).
 (c) Full-angle distributions at the same Coulomb logarithm as the
 distribution broadens toward isotropy.}
 \label{fig:app-moliere-width}
\end{figure}

The Moli\`ere width \(\omega\) in Eq.~\eqref{eq:moliere-width} is derived from
the binary cross section rather than fitted to the cumulative distribution.
Figure~\ref{fig:app-moliere-width} tests this scale as the initially narrow
forward population develops and eventually broadens over the sphere.

Throughout most of the broad range \(\omega<\pi/4\), the width predicted by
Eq.~\eqref{eq:moliere-width} differs from the standard deviation fitted to
direct collision-by-collision Monte Carlo distributions by less than \(2\%\)
[Fig.~\ref{fig:app-moliere-width}(a)].  The nearly common slopes in panel (b)
confirm the predicted Gaussian scale, while the higher Laguerre terms in
Eq.~\eqref{eq:laguerre-core} describe the remaining shape.  At \(N=O(1)\), a
smooth cumulative core has not yet formed.  By the time
\(\omega\sim\pi/4\), the distribution has generally entered the
low-order relaxation regime and is already broadening toward isotropy
[Fig.~\ref{fig:app-moliere-width}(c)].

\subsection{Uniform core--tail expansion}

For \(N\gg1\), and hence \(B\gg1\), the Hankel exponent can be reorganized as a
controlled expansion in \(1/B\) that retains both the forward and large-\(z\)
limits.

The \(1/B\) structure below is akin to the classical Moli\`ere
expansion, since both reflect the same logarithmic small-angle behavior of the
Coulomb kernel~\cite{bethe_molieres_1953}. Moli\`ere theory, however,
constructs this expansion after reducing the collision geometry to its
small-angle form (assuming $\cos\chi\sim1-\chi^2/2,~\sin\chi\sim\chi$), although 
it also discusses how the large-angle part of the Moli\`ere distribution becomes 
dominated by a single collision.

Here the expansion is recovered from the rescaled GS
representation, and its \(k=\Lambda\chi\to\infty\) limit explicitly connects
the forward core to the \(\chi^{-4}\) tail and its cumulative correction within
the same Boltzmann coefficient hierarchy.

Let \(q=\omega\rho\) and \(z=\chi^2/(2\omega^2)\).  Using the small-\(x\)
form of Eq.~\eqref{eq:rescaled-core-branch} together with
\(\omega^2=2NB/\Lambda^2\), the exponent of the Hankel kernel becomes
\begin{equation}
 N[\tilde{\varsigma}(x)-1]
 =-\frac{q^2}{2}
 +\frac{q^2}{B}\ln\frac{q}{\sqrt{2}}
 -\frac{q^4}{16NB^2}+o(B^{-2}).
 \label{eq:app-uniform-exponent}
\end{equation}
Expanding the Hankel integral in \(1/B\) gives
\begin{equation}
\begin{split}
 g(\chi)\simeq{}&
 \sqrt{\frac{\chi}{\sin\chi}}\frac{1}{2\pi\omega^2}
 \left[e^{-z}+\frac{F_1(z)}{B}\right.\\
 &\left. +\frac{1}{B^2}\left\{F_2(z)
 -\frac{e^{-z}}{4N}(z^2-4z+2)\right\}\right],
\end{split}
 \label{eq:app-uniform-expansion}
\end{equation}
where
\begin{align}
 F_1(z)&=(z-1)e^{-z}
 [\operatorname{Ei}(z)-\ln z]+2e^{-z}-1,
 \label{eq:app-uniform-functions}\\
 F_2(z)&=
 \frac12\int_0^\infty v^2(\ln v)^2e^{-v}
 J_0(2\sqrt{zv})\,\dd v.\nonumber
\end{align}
The first term is the Gaussian core.  At large \(z\),
\begin{align}
 F_1(z)&\sim z^{-2}+4z^{-3}+\cdots,\nonumber\\
 F_2(z)&\sim
 (4\ln z+8\gamma_E-12)z^{-3}+\cdots ,
 \label{eq:app-uniform-large-z}
\end{align}
so the same expansion recovers the \(\chi^{-4}\) law and the
\(\chi^{-6}\ln\chi\) correction in Eq.~\eqref{eq:analytic-tail}.  The second
tail order combines the subleading part of \(F_1\) with the leading part of
\(F_2\), showing that the core and tail arise from the same expansion.

The non-elementary \(F_2\) makes neither the cumulative distribution nor its
inverse explicit.  Equation~\eqref{eq:app-uniform-expansion} therefore
establishes the analytic connection between the two angular limits, whereas
the practical representation uses the explicit Laguerre core and the
optimally truncated second-order tail tested in
Figs.~\ref{fig:core-tail-emergence}(b) and~\ref{fig:matching-regimes}(a).

\section{Sampling Representation}
\label{app:sampling}

The three statistical regimes identified in Sec.~\ref{sec:statistical-regimes}
require different representations of the same GS distribution but share one
sampling rule.  Let \(g_r(\chi)\) denote the smooth scattered density in the
representation appropriate to the current \(N\) and \(\Lambda\).  Since its
total weight is \(1-e^{-N}\), the conditional cumulative distribution is
\begin{equation}
 {\cal F}_r(\chi)=
 \frac{2\pi}{1-e^{-N}}
 \int_0^\chi g_r(\chi')\sin\chi'\,\dd\chi'.
 \label{eq:app-conditional-cdf}
\end{equation}
For a uniform random number \(U\), \(U<e^{-N}\) returns the unscattered state
\(\chi=0\).  Otherwise,
\begin{equation}
 \chi={\cal F}_r^{-1}\!
 \left(\frac{U-e^{-N}}{1-e^{-N}}\right)
 \label{eq:app-inverse-sampling}
\end{equation}
samples the scattered population.

The choice of \(g_r\) follows directly from the regime structure.  For
\(N<N_{\min}=O(1)\), a smooth cumulative core is absent: an integer
\(K\sim{\rm Poisson}(N)\) is drawn, and \(K\) independent angles from the
normalized binary cross section are composed successively.  At intermediate
\(N\), \(g_r=g_{\rm comp}\) from Eq.~\eqref{eq:composite-distribution}, whose
core and tail CDFs meet at the intrinsic \(\chi_c\).  When
\((L+1/2)e^{-s_L}\leq\delta_L\), only a few GS coefficients remain appreciable
and \(g_r\) is instead evaluated from the series truncated at \(L\).  Its CDF
follows term by term, for \(l\geq1\), from
\begin{equation}
 \int_0^\chi P_l(\cos\chi')\sin\chi'\,\dd\chi'
 =\frac{P_{l-1}(\cos\chi)-P_{l+1}(\cos\chi)}{2l+1}.
 \label{eq:app-legendre-cdf}
\end{equation}
Thus the switches select numerically efficient representations rather than
different collision models.  The practical value of \(N_{\min}\), the retained
Laguerre and Legendre orders, the explicit branch CDFs, and the inversion
procedure are given in the SM, Sec. II.~\cite{SM}.

After \(\chi\) is obtained, an azimuth is drawn uniformly on
\([0,2\pi)\).  The relative momentum is rotated by these two angles in the
center-of-momentum frame and transformed back to the laboratory frame.  This is the
standard relativistic binary-collision update. ACES modifies only the sampled
angular distribution.  Collision pairing, relativistic transformations, and the conservative
update for unequally weighted macroparticles are otherwise unchanged
\cite{perez_improved_2012,higginson_corrected_2020}.  The corresponding
pseudocode, saddle-point evaluation of the Laguerre moments, closed-form CDFs,
and full-range sampling tests are deferred to the Supplemental
Material~\cite{SM}.

\section{Simulation and Diagnostic Definitions}
\label{app:simulation-details}

Both applications in Sec.~\ref{sec:physical-consequences} were calculated
with the two-dimensional relativistic PIC code KLAPS \cite{wang_integrated_2015}.
The Maxwell solver and ionization were disabled. The particle-pusher time step satisfies
$c\Delta T=0.64\min(\Delta x,\Delta y)$, and collisions are subcycled with
$\Delta t_{\rm coll}=0.2\Delta T$.  Unequally weighted macroparticles are
updated with the conservative procedure of Ref.~\cite{higginson_corrected_2020}.
The ACES and NP calculations use identical grids, particle loading, pairing,
and relativistic kinematics, and only the sampled scattering-angle distribution is
changed.  A separate long-time electron--ion relaxation benchmark shows that
the ACES and NP electron and ion energy histories remain nearly
indistinguishable over $62{,}500$ time steps, confirming the same mean
interspecies energy relaxation over repeated updates. The comparison is given
in the SM, Sec. IV~\cite{SM}. The principal parameters are summarized in
Table~\ref{tab:simulation-parameters}.

For the fast-electron calculation, a uniform monoenergetic beam with kinetic
energy $2$ MeV and transverse half-width $3~\mu\mathrm m$ is continuously
injected along $+x$. The initial transverse momentum is zero. Forward-moving
beam electrons with kinetic energy above $50$ keV are recorded once at each
crossing plane, with adjacent planes separated by $5~\mu\mathrm m$. The
deflection angle in Fig.~\ref{fig:electron-transport}(c) is 
$\chi=\arctan\!\left(\sqrt{p_y^2+p_z^2}/{p_x}\right),~p_x>0$.

The spectra and transverse profiles in Fig.~\ref{fig:electron-transport} are
accumulated over the full injection interval.  Each crossing contributes its
macroparticle weight to the energy and angular histograms and
$w(\gamma-1)m_ec^2$ to the kinetic-energy flux.  The latter weight is also used
for the aperture-integrated quantity in Eq.~\eqref{eq:aperture-energy-flux}.

\begin{table}[ht]
\caption{Parameters of the fast-electron and alpha-particle calculations.
Here O and P denote open and periodic boundaries, respectively, and PPC gives
the macroparticles per cell in the order $(e,{\rm DT},\alpha)$.}
\label{tab:simulation-parameters}
\begin{ruledtabular}
\setlength{\tabcolsep}{3pt}
\begin{tabular}{lcc}
 & Fast electron & Alpha particle \\
\colrule
Domain $(\mu\mathrm m^2)$ & $64\times48$ & $16\times4$ \\
Grid & $2048\times1536$ & $1024\times256$ \\
Boundaries $(x,y)$ & O, P & P, P \\
$T_e=T_{\rm DT}$ & $1$ keV & $5$ keV \\
$n_e=n_{\rm DT}$ & $2.2\times10^{25}~\mathrm{cm}^{-3}$
 & $1.1\times10^{25}~\mathrm{cm}^{-3}$ \\
PPC $(e,{\rm DT},\alpha)$ & $(36,36,0)$ & $(36,36,16)$ \\
\end{tabular}
\end{ruledtabular}
\end{table}

For the alpha-particle calculation, an isotropic $3.54$-MeV alpha population
with $n_\alpha=0.05n_{\rm DT}$ is initialized uniformly in the periodic domain.
The DT-ion and alpha-particle energy distributions in
Fig.~\ref{fig:alpha-particle} are macroparticle-weighted histograms over the
entire domain.  All collision pairs and numerical parameters are held fixed
between the ACES and NP calculations, so their difference measures only the
effect of the cumulative scattering statistics.

\bibliography{collision}

\end{document}